\ifdefined\OSMSeparateAppendix
  \documentclass[article,shortnames]{jss}
\else
  \documentclass[article,shortnames,nojss]{jss}
\fi

\newif\ifseparateappendix
\ifdefined\OSMSeparateAppendix \separateappendixtrue \fi

\usepackage{orcidlink,thumbpdf,lmodern}
\usepackage{amsmath,amssymb,bm,booktabs,array}
\usepackage{needspace}   %
\makeatletter
\renewcommand{\jsssec}[2][default]{\vskip \preXLskip%
  \pdfbookmark[1]{#1}{Section.\thesection.#1}%
  \refstepcounter{section}%
  \centerline{\textbf{\Large \thesection. #2}} \nopagebreak
  \vskip \postMskip \nopagebreak \@afterheading}
\renewcommand{\jsssecnn}[1]{\vskip \preXLskip%
  \centerline{\textbf{\Large #1}} \nopagebreak
  \vskip \postMskip \nopagebreak \@afterheading}
\renewcommand{\jsssubsec}[2][default]{\vskip \preMskip%
  \pdfbookmark[2]{#1}{Subsection.\thesubsection.#1}%
  \refstepcounter{subsection}%
  \textbf{\large \thesubsection. #2} \nopagebreak
  \vskip \postSskip \nopagebreak \@afterheading}
\renewcommand{\jsssubsecnn}[1]{\vskip \preMskip%
  \textbf{\large #1} \nopagebreak
  \vskip \postSskip \nopagebreak \@afterheading}
\renewcommand{\jsssubsubsec}[2][default]{\vskip \preSskip%
  \pdfbookmark[3]{#1}{Subsubsection.\thesubsubsection.#1}%
  \refstepcounter{subsubsection}%
  {\large \textit{#2}} \nopagebreak
  \vskip \postSskip \nopagebreak \@afterheading}
\renewcommand{\jsssubsubsecnn}[1]{\vskip \preSskip%
  {\textit{\large #1}} \nopagebreak
  \vskip \postSskip \nopagebreak \@afterheading}
\AddToHook{para/before}{\if@nobreak\nobreak\fi}   %
\let\osm@makefooter\makefooter
\renewcommand{\makefooter}{\par\Needspace*{16\baselineskip}\osm@makefooter}
\makeatother
\newcommand{\Cov}{\operatorname{Cov}}
\newcommand{\Var}{\operatorname{Var}}
\newcommand{\Ex}{\operatorname{E}}
\newcommand{\bB}{\bm{B}}
\newcommand{\balpha}{\bm{\alpha}}
\newcommand{\bw}{\bm{w}}
\newcommand{\countword}[1]{\ifcase#1\relax zero\or one\or two\or three\or four\or
  five\or six\or seven\or eight\or nine\or ten\else #1\fi}

\providecommand{\atClosedSizes}{114}
\providecommand{\atClosedSizeMax}{1001}

\providecommand{\atClosedWorst}{$5.33 \times 10^{-15}$}
\providecommand{\atClosedWorstN}{801}
\providecommand{\atClosedWorstFamily}{exponential}
\providecommand{\atClosedWorstQuantity}{mean}

\providecommand{\atClosedCovEntries}{2240}
\providecommand{\atClosedQuadErrMax}{$2.37 \times 10^{-10}$}
\providecommand{\atAnchorLogis}{$0$}
\providecommand{\atAnchorGumbel}{$5.55 \times 10^{-17}$}

\providecommand{\atIdentSizesNorm}{99}
\providecommand{\atIdentOtherSizes}{5, 10, 25, 50, 100}

\providecommand{\atIdentRowSumsWorst}{$1.93 \times 10^{-13}$}
\providecommand{\atIdentReflWorst}{$3.55 \times 10^{-15}$}
\providecommand{\atIdentWorstAll}{$9.95 \times 10^{-13}$}
\providecommand{\atIdentWorstAllFamily}{exponential}
\providecommand{\atIdentWorstAllN}{100}

\providecommand{\atIdentTimeHundredMin}{3.0}

\providecommand{\atLargeN}{1001}

\providecommand{\atLargeCores}{12}
\providecommand{\atLargePairs}{250,500}
\providecommand{\atLargeEntries}{500,500}

\providecommand{\atLargeWallMin}{47.0}
\providecommand{\atLargeRowMax}{$5.72 \times 10^{-12}$}
\providecommand{\atLargeRowMean}{$3.57 \times 10^{-14}$}
\providecommand{\atLargeTotal}{$3.14 \times 10^{-11}$}
\providecommand{\atLargeSumAlpha}{$1.64 \times 10^{-14}$}
\providecommand{\atLargeSumSq}{$1.36 \times 10^{-12}$}

\providecommand{\atTimeReps}{3}

\providecommand{\atTimeNmin}{5}
\providecommand{\atTimeNmax}{1001}

\providecommand{\atTimeSingles}{15}
\providecommand{\atTimeDoublesPkg}{6}

\providecommand{\atTimePerSingleAtMin}{0.26}
\providecommand{\atTimePerSingleAtMax}{0.28}
\providecommand{\atTimePerDoubleAtMin}{43.5}
\providecommand{\atTimePerDoubleAtMax}{94.3}

\providecommand{\atTimeShareDoubleMin}{99}
\providecommand{\atTimeShareDoubleMax}{99}

\providecommand{\atTimeMachine}{Apple M3 Pro}
\providecommand{\atFigZeroCount}{123}

\providecommand{\atClosedMargin}{18765}

\providecommand{\atQuadErrRelation}{every residual stays below the estimate of its own call}

\providecommand{\atClosedSizesAboveHundred}{15}
\providecommand{\atTrendClosed}{grows by a factor of at most 6.0 (exponential)}
\providecommand{\atTimeDoubleGrowth}{2.2}

\providecommand{\atLargeNEq}{$n = 10^{4}$, $10^{5}$, and $10^{6}$}
\providecommand{\atLargeNMax}{10^{6}}

\providecommand{\atLargeNMeanWorst}{$1.11 \times 10^{-10}$}
\providecommand{\atLargeNVarWorst}{$7.99 \times 10^{-11}$}

\providecommand{\atLargeNCorrWorst}{$3.47 \times 10^{-11}$}

\providecommand{\atLargeNExtDigits}{5 to 10}

\providecommand{\atLargeNOtherDigits}{8}

\providecommand{\atLargeNTimeCovMin}{0.8}
\providecommand{\atLargeNTimeCovMax}{1.0}
\providecommand{\atLargeNPerDoubleMs}{98}

\providecommand{\cpTietjenDoubtful}{1}
\providecommand{\cpTietjenFail}{33}
\providecommand{\cpTietjenFailNmin}{29}
\providecommand{\cpTietjenWorstFail}{$1.37 \times 10^{-4}$}

\providecommand{\cpTietjenUnitsMax}{1372}
\providecommand{\cpOtherUnitsMax}{3.4}

\providecommand{\cpLiebleinWorstMean}{$4.12 \times 10^{-7}$}
\providecommand{\cpLiebleinWorstCov}{$4.38 \times 10^{-6}$}

\providecommand{\cpGuptaFiveUnits}{2.1}
\providecommand{\cpGuptaTenFail}{13}
\providecommand{\cpGuptaTenCells}{20}
\providecommand{\cpGuptaTenFailRankMin}{4}

\providecommand{\cpLogisWorst}{$8.88 \times 10^{-16}$}
\providecommand{\cpLogisCells}{7}

\providecommand{\cpEnvStatsWorst}{$1.13 \times 10^{-13}$}
\providecommand{\cpSuppDistsWorst}{$7.92 \times 10^{-5}$}

\providecommand{\cpLmomcoNote}{fails with a non-finite function value at location 0 and runs at location 1, so the normal was called at location 1 and shifted back}
\providecommand{\cpMosExactWorst}{$1.72 \times 10^{56}$}
\providecommand{\cpMosMcMeanRel}{0.19}
\providecommand{\cpMosMcVarRel}{1.0}

\providecommand{\cpMetaBlueAlphaRel}{0.31}
\providecommand{\cpMetaBlueCovRel}{6.0}
\providecommand{\cpMetaBlueExtVarRelAtHundredOne}{2.1}
\providecommand{\cpMetaBlueExtVarRelAtThousand}{3.1}
\providecommand{\cpMetaBlueExtVarRelAtFive}{6.0}

\providecommand{\cpToolVersions}{\pkg{EnvStats} 3.1.0, \pkg{SuppDists} 1.1.9.9, \pkg{lmomco} 2.5.8, \pkg{mos} 0.1.4, \pkg{metaBLUE} 1.0.0}
\providecommand{\cpScipyVersion}{1.18.1}
\providecommand{\cpPythonVersion}{3.15.2}
\providecommand{\cpNumpyVersion}{2.5.4}

\providecommand{\cpScipyEntries}{250}

\providecommand{\cpScipyWorstMean}{$6.41 \times 10^{-10}$}
\providecommand{\cpScipyWorstVar}{$1.67 \times 10^{-13}$}

\providecommand{\cpCellsTotal}{556}

\providecommand{\cpPublicationsTotal}{7}
\providecommand{\cpCellsAgree}{507}
\providecommand{\cpCellsTable}{49}

\providecommand{\cpCellsFine}{22}

\providecommand{\cpRouteMaxDiffAll}{$2.55 \times 10^{-12}$}
\providecommand{\cpFineUnitsMax}{9.6}
\providecommand{\cpTietjenGross}{20}
\providecommand{\cpTietjenFine}{13}
\providecommand{\cpTietjenFineUnitsMax}{9.1}
\providecommand{\cpGuptaFiveAgree}{9}
\providecommand{\cpGuptaFiveFail}{1}
\providecommand{\cpGuptaTenUnitsMin}{1.7}
\providecommand{\cpLmomcoSoundTo}{101}
\providecommand{\cpLmomcoSoundWorst}{$9.77 \times 10^{-7}$}
\providecommand{\cpLmomcoFirstFail}{501}
\providecommand{\cpMosExpSoundTo}{25}
\providecommand{\cpMosExpSoundWorst}{$1.79 \times 10^{-7}$}
\providecommand{\cpMosExpFirstFail}{101}
\providecommand{\cpMosUnifWorst}{$1.97 \times 10^{-13}$}
\providecommand{\cpEnvStatsSoundTo}{1001}

\providecommand{\cpExactCells}{30}
\providecommand{\cpExactPkgMax}{$4.88 \times 10^{-15}$}

\providecommand{\cpExactTenFail}{13}

\providecommand{\cpExactTenMax}{$3.92 \times 10^{-4}$}
\providecommand{\cpExactTenUnitsMax}{392}

\providecommand{\fmScipyVersion}{1.18.1}
\providecommand{\fmTolScipy}{10^{-8}}

\providecommand{\fmRefValidCells}{52}
\providecommand{\fmRefValidWorstMean}{8.67 \times 10^{-14}}
\providecommand{\fmRefValidWorstVar}{1.18 \times 10^{-10}}

\providecommand{\fmCheckCases}{19}

\providecommand{\fmCheckValues}{926}

\providecommand{\fmCheckPkgRefWorst}{1.47 \times 10^{-13}}

\providecommand{\fmCheckScipyWorstCase}{\code{beta(shape1 = 0.5, shape2 = 0.5)}}
\providecommand{\fmCheckScipyWorstN}{1001}
\providecommand{\fmCheckScipyWorstR}{1001}
\providecommand{\fmCheckScipyOver}{20}
\providecommand{\fmCheckScipyOverWhere}{all of them at the extreme rank next to an endpoint where the density is infinite}
\providecommand{\fmCheckScipyWorstPct}{20}

\providecommand{\fmIdentMatrices}{36}

\providecommand{\fmIdentWorst}{2.54 \times 10^{-12}}
\providecommand{\fmCovPairs}{161}
\providecommand{\fmCovRouteFailed}{22}
\providecommand{\fmCovN}{25}
\providecommand{\fmCovWorst}{3.10 \times 10^{-10}}

\providecommand{\fmScipyCauchyMinMean}{-56.6}
\providecommand{\fmScipyCauchyMinVar}{1.18 \times 10^{16}}
\providecommand{\fmScipyCauchySecondVar}{71.0}
\providecommand{\fmScipyNonexistCells}{6}

\providecommand{\fmScipyNonexistWarnPhrase}{without a warning}
\providecommand{\fmScipyTMinVar}{44.25}
\providecommand{\fmRefTMinVar}{45.69}
\providecommand{\fmScipyTMinPct}{3.2}
\providecommand{\fmPkgTMinDiff}{4.50 \times 10^{-14}}
\providecommand{\fmScipyArcsinePct}{20}
\providecommand{\fmPkgArcsineDiff}{6.66 \times 10^{-16}}
\providecommand{\fmRegionReturned}{860}

\providecommand{\fmRegionErrors}{4}
\providecommand{\fmRegionCells}{864}
\providecommand{\fmRegionErrorList}{\code{weibull(shape = 0.2)} at rank $n$ of $n = 1001$, $10^{4} + 1$, $10^{6} + 1$, and \code{f(df1 = 4, df2 = 4.02)} at rank $n$ of $n = 10^{6} + 1$}
\providecommand{\fmRegionCases}{30}
\providecommand{\fmRegionPkgWorst}{4.37 \times 10^{-11}}
\providecommand{\fmRegionScipyOver}{175}
\providecommand{\fmRegionScipyNonFinite}{180}
\providecommand{\fmRegionScipyNonFiniteFrom}{10^{4} + 1}
\providecommand{\fmRegionScipyFinite}{684}
\providecommand{\fmMatComputed}{25}
\providecommand{\fmMatErrors}{6}
\providecommand{\fmMatCases}{34}
\providecommand{\fmMatN}{20}
\providecommand{\fmMatErrorCases}{\code{gamma(shape = 0.05)}, \code{weibull(shape = 0.2)}, \code{t(df = 2.1)}, \code{t(df = 2.3)}, \code{t(df = 2.05)}, \code{t(df = 2.2)}}
\providecommand{\fmMatNotApplicable}{3}
\providecommand{\fmMatWorstSumB}{9.97 \times 10^{-10}}
\providecommand{\rpFiveN}{25}
\providecommand{\rpFiveSeMeanFactor}{0.2069}
\providecommand{\rpFiveSeSdFactor}{0.1586}
\providecommand{\rpFiveSeSampleMean}{0.2}
\providecommand{\rpFiveSeMeanPct}{3.4}
\providecommand{\rpFiveWLocExtreme}{0.0821}
\providecommand{\rpFiveWLocMedian}{0.2644}
\providecommand{\rpLogisN}{9}
\providecommand{\rpLogisVarLoc}{0.3425}
\providecommand{\rpLogisVarScale}{0.0863}
\providecommand{\rpLogisSdPerScale}{1.8138}
\providecommand{\rpMedTenRanks}{5 and 6}
\providecommand{\rpMedTenWeight}{0.50}
\providecommand{\rpMedTenVar}{0.13833}
\providecommand{\rpMedElevenVar}{0.13716}
\providecommand{\rpMedRatio}{1.008}
\providecommand{\rpPkgVersion}{0.3.0}

\providecommand{\rpRVersion}{R version 4.6.2 (2026-10-29)}
\providecommand{\rpUnsplitN}{10^{6}}
\providecommand{\rpUnsplitMass}{8.8 \times 10^{-35}}
\providecommand{\rpUnsplitErr}{1.6 \times 10^{-34}}

\providecommand{\rpCenterSqRatioMax}{0.21}

\providecommand{\rpLognormOuterReach}{1.5 \times 10^{7}}
\providecommand{\rpMixMass}{0.9914}
\providecommand{\rpLaplaceSumBExact}{10}
\providecommand{\rpLaplaceNoBreaksSumB}{1.7 \times 10^{-8}}
\providecommand{\rpLaplaceNoBreaksDiff}{4.3 \times 10^{-9}}
\providecommand{\rpLaplaceNoBreaksPair}{1 and 2}
  \ifseparateappendix  \fi   %

\author{Zheyang Wu\\Worcester Polytechnic Institute
   \And Tiejun Tong\\Hong Kong Baptist University}
\Plainauthor{Zheyang Wu, Tiejun Tong}

\title{\pkg{orderStatMoments}: Exact Means, Variances, and Covariances of Order Statistics in \proglang{R}}
\Plaintitle{orderStatMoments: Exact Means, Variances, and Covariances of Order Statistics in R}
\Shorttitle{\pkg{orderStatMoments}: Moments of Order Statistics in \proglang{R}}

\Abstract{
  Calculating the moments, i.e., means, variances, and covariances, of the order statistics
  of a sample is critical, as they are foundations for order-statistics-based
  estimation and inference problems. In the early literature, the exact covariance
  matrices were tabulated only up to the sample size of 50 under normality. Recent software
  either provides exact calculation only for the means and variances of single order
  statistics, or obtains the covariances by approximations.

  We create the \proglang{R} package \pkg{orderStatMoments} to calculate the exact values of
  these moments for any ranks and any sample size. It covers thirteen built-in distribution
  families, specified by name and parameters as in \proglang{R}, and allows for continuous
  distributions defined by the user under relatively weak conditions. It also calculates
  the moments of any linear combination of order statistics, including sample quantiles,
  and Lloyd's best linear unbiased estimators with their variances.

  The work makes two contributions. First, the package provides the exact covariances
  between order statistics, which no software we found does. Second, we design methods based on the
  statistical properties of order statistics critical for ultra accurate computing at large
  sample sizes and various distributions. The accuracy is
  verified against closed forms at sample sizes up to $n = \atLargeNMax$, against the
  permutation identities on full matrices, and against \cpCellsTotal{} cells tabulated in
  \countword{\cpPublicationsTotal} publications, where it identified some errors far beyond the
  stated accuracy. A comparison with five \proglang{R} packages and
  \proglang{Python}'s \pkg{SciPy} shows advantages of this package. Computing is
  efficient, with a time nearly independent of the sample size.
}

\Plainkeywords{order statistics, moments, numerical integration, best linear unbiased estimator, sample quantiles, five-number summary, R}

\Address{
  Zheyang Wu\\
  Department of Mathematical Sciences\\
  Worcester Polytechnic Institute\\
  Worcester, MA, United States of America\\
  E-mail: \email{zheyangwu@wpi.edu}\\[1em]
  Tiejun Tong\\
  Department of Mathematics\\
  Hong Kong Baptist University\\
  Hong Kong, China\\
  E-mail: \email{tongt@hkbu.edu.hk}
}

\begin{document}

\section{Introduction} \label{sec:intro}

To compute the uncertainty of a statistic built from ordered observations, one often needs
to calculate their variances and covariances. Order statistics are the sorted values
$X_{(1)} \le \dots \le X_{(n)}$ of a random sample of $n$ independent draws from the same
distribution, with  
\[
\text{means } \mu_{r:n} = \Ex[X_{(r)}] \quad \text{and covariances }
\sigma_{r,s:n} = \Cov(X_{(r)}, X_{(s)}).
\] 
Calculation of these moments is important. For example, the best linear unbiased
estimators (BLUE) of location and scale of \citet{lloyd1952least} are built from the mean
vector and the covariance matrix of the order statistics. Calculating the mean and the
variance of any linear combination of order statistics, including a trimmed mean, a range,
or a sample quantile, also requires the mean vector and the covariance matrix of the order
statistics. Moreover, in meta-analysis, studies may report a median, quartiles, or a
five-number summary (the minimum, the three quartiles, and the maximum) in place
of the sample mean and sample standard deviation. Several
methods have been proposed to impute the mean and standard deviation from these order
statistics \citep{hozo2005estimating, wan2014estimating, luo2018optimally,
shi2020estimating, balakrishnan2022unified}. Calculating the variances of these imputed
values requires the means, variances, and covariances of the order statistics involved.

The exact moments of the order statistics were first made available as printed tables, which
cover small samples only. Under the normal distribution, \citet{teichroew1956tables} and
\citet{sarhan1956estimation} tabulated the means, variances, and covariances for $n \le 20$,
and \citet{tietjen1976variances} and \citet{parrish1992expected, parrish1992computing}
extended them to $n \le 50$. Beyond $n = 50$, only the means \citep{harter1961expected} and
the mean and variance of the range \citep{harter1960tables} were tabulated. For other
distributions the tables are smaller. \citet{gupta1960order} gave the moments under the
gamma distribution with integer shape up to $n = 15$, \citet{gupta1965exact} tabulated the
first four moments under the logistic distribution up to $n = 10$, and
\citet{chen1997moments} provided the moments under the inverse Gaussian distribution up to
$n = 25$. Overall, no printed covariance table we found reaches $n > 50$ for the normal
distribution or $n > 25$ for other distributions.

The relatively recent software we found covers the exact calculation of the means and
variances, but approximates the covariances. For the means of normal order statistics,
\pkg{EnvStats} \citep{millard2025envstats} and \pkg{SuppDists}
\citep{wheeler2025suppdists} implement the exact form and the approximate form of
Algorithm AS~177 of \citet{royston1982algorithm}, respectively. For one order statistic of
a general distribution, \pkg{lmomco} \citep{asquith2026lmomco} returns the mean by
quadrature, \pkg{mos} \citep{arafeh2026mos} returns the mean and variance by closed forms
or by Monte Carlo, and the function \code{scipy.stats.order\_statistic} of \pkg{SciPy}
\citep{virtanen2020scipy} computes the mean and variance by numerical integration. For the
covariances, the tools we found rely on the Taylor series of
\citet{david1954statistical}: \pkg{metaBLUE} \citep{yang2018metablue} for normal order
statistics, Algorithm AS~128 \citep{davis1978algorithm}, which is offered commercially by
the Numerical Algorithms Group (NAG), and the \proglang{Maple} procedures of
\citet{childs2002series}. None of these tools computes the covariances between order
statistics by exact integration, for the normal or other distributions.

To fill this gap, we implement a new \proglang{R} package, \pkg{orderStatMoments}. It
computes the exact integrals that define the means, variances, and covariances by adaptive
quadrature, for any ranks and any sample size. It covers thirteen built-in families: the
twelve continuous distributions for which base \proglang{R} provides the distribution,
density, and quantile functions (the normal, uniform, exponential, logistic,
Student's $t$, Cauchy, gamma, chi-square, beta, lognormal, Weibull, and $F$ distributions)
and the Gumbel distribution. A built-in family is specified as in \proglang{R}, by its name
and its parameters. The package also covers a continuous
distribution that the user supplies through its distribution, density, and quantile
functions, provided the distribution function and the density can be evaluated on the log
scale in both tails and the requested moments exist (Section~\ref{sec:userdist}). Where a
built-in family has a closed-form expression for a moment, the package returns the
closed-form value in place of the quadrature. From these moments, the package also computes
the mean and the variance of any linear combination of order statistics, such as a sample
quantile, and Lloyd's BLUE together with their variances. Throughout, \emph{exact} means the
defining integral computed numerically to a stated tolerance, as opposed to a series
approximation or a simulation.

We make two main contributions. The first is the exact calculation of the covariances. The
package computes each covariance from its defining double integral, for any pair of ranks,
any sample size, and any distribution it covers, which none of the tools above does.

The second contribution is a computing methodology that we design from the statistical
properties of order statistics, in place of a direct call to a general-purpose integrator.
A direct call faces several difficulties. For example, at large $n$, the density of an order
statistic is a narrow peak that a typical adaptive rule can miss. Also, the coefficients and
the tail probabilities of the integrand often overflow or underflow, and a variance that is
formed as the difference of two moments can lose its accuracy by cancellation. Moreover, a
moment that does not exist can come back as a finite number. Our design addresses these
difficulties as one system. First, the distribution theory of
order statistics gives the center and the width of each order statistic before any
integration. We use them to take each moment about the
center of its own order statistic, to cut the integration range around the peak, and to
scale the tolerances. Second, the tails of the parent distribution decide at which ranks a
moment exists, and the package refuses a moment that does not exist. Third, the package
checks its results against facts that hold for order statistics, such as a density that
integrates to one.

Two examples show why such a design matters. The function
\code{scipy.stats.order\_statistic} of \pkg{SciPy} (version \fmScipyVersion) computes the
mean and the variance of one order statistic by numerical integration. For the minimum of
five Cauchy observations, whose mean and variance do not exist, it returns a mean of
$\fmScipyCauchyMinMean$ and a variance of $\fmScipyCauchyMinVar$ without a warning, where
our package stops with an error that names the rank. For the minimum of five observations
from Student's $t$ distribution with 2.1 degrees of freedom, whose variance exists, it
returns a variance of \fmScipyTMinVar, which is \fmScipyTMinPct\% away from the accurate
value \fmRefTMinVar{}. This value comes from the independent reference integral of
Section~\ref{sec:newfamilies}, and our package agrees with it within a relative difference
of $\fmPkgTMinDiff$.

The paper is organized as follows. Section~\ref{sec:computation} presents the methodology:
the target problem, the related challenges and our solutions to them, the closed-form
results, and the package's internal checks. Section~\ref{sec:package} introduces the
package usage, with two illustrations. Section~\ref{sec:accuracy} lays out
the evidence of its accuracy. Section~\ref{sec:comparison} compares the package with the
other tools and reports the computing time. Section~\ref{sec:summary} summarizes the
results and discusses the limitations of the package. The main text presents the main
ideas, and \ifseparateappendix the technical appendix, a separate supplementary file, gives
\else the appendices give \fi the further details under similar titles.

\section{The exact moments and their computation} \label{sec:computation}

The integral definitions of the moments of order statistics are well known, but computing
them, especially at large $n$, is relatively challenging. Section~\ref{sec:integrals} gives
the integrals and the linear algebra built on them. Section~\ref{sec:background} explains
why a direct numerical integration fails, and Section~\ref{sec:devices} presents our
solutions. Section~\ref{sec:closed} lists the closed forms used in place of the quadrature,
the condition under which a moment exists, and the package's internal checks against
quadrature errors.

\subsection{The integrals and the linear statistics} \label{sec:integrals}

The exact moment calculations are based on integrals. Let $F$ be the cumulative distribution
function, $f$ be the density, and $Q = F^{-1}$ be the quantile function of the continuous
parent distribution from which the sample is drawn. The density of $X_{(r)}$ and the joint
density of $X_{(r)}$ and $X_{(s)}$ for $r < s$ are \citep[Chapter~2]{david2003order}
\begin{align}
f_{r:n}(x) &= \frac{n!}{(r-1)!\,(n-r)!}\,F(x)^{r-1}\{1 - F(x)\}^{n-r} f(x),
\label{eq:single}\\
f_{r,s:n}(y, z) &= \frac{n!}{(r-1)!\,(s-r-1)!\,(n-s)!}\,
F(y)^{r-1}\{F(z) - F(y)\}^{s-r-1} \nonumber\\
&\qquad \times \{1 - F(z)\}^{n-s} f(y) f(z), \qquad y < z.
\label{eq:double}
\end{align}
The moments are the integrals against these densities over the support of the parent
distribution,
\begin{equation}
\begin{aligned}
\mu_{r:n} &= \int x\, f_{r:n}(x)\, dx, \qquad
\Ex[X_{(r)}^2] = \int x^2 f_{r:n}(x)\, dx, \\
\Ex[X_{(r)} X_{(s)}] &= \iint_{y < z} yz\, f_{r,s:n}(y, z)\, dy\, dz,
\end{aligned}
\label{eq:moments}
\end{equation}
and the variances and covariances follow as $\sigma_{r,r:n} = \Ex[X_{(r)}^2] -
\mu_{r:n}^2$ and $\sigma_{r,s:n} = \Ex[X_{(r)} X_{(s)}] - \mu_{r:n}\mu_{s:n}$.

The package computes these integrals for \emph{parent} distributions of $X$, including
thirteen built-in families and a distribution the user supplies through its $F$, $f$, and $Q$
(any continuous distribution that meets the weak conditions of
Section~\ref{sec:userdist}). It also allows a straightforward extension to the moments of
the shifted and scaled variable $Y = a + bX$ with $b > 0$ (via the function arguments
\code{os_shift} $= a$ and \code{os_multiplier} $= b$):
\begin{equation*}
\Ex[Y_{(r)}] = a + b\,\mu_{r:n}, \qquad \Cov(Y_{(r)}, Y_{(s)}) = b^2\,\sigma_{r,s:n}.
\end{equation*}
Here the shift $a$ and the multiplier $b$ are known
constants that the user gives. They are not to be confused with the unknown location and
scale parameters $\mu$ and $\sigma$ that Lloyd's BLUE in \eqref{eq:blue} estimates. Our
methodology discussion is based on $X$ by default, unless specified otherwise.

Moment calculations are critical to linear estimators based on the order statistics. For a
rank set $r_1 < \dots < r_k$ we write $\bm X = (X_{(r_1)}, \dots, X_{(r_k)})'$,
$\balpha = \Ex[\bm X]$ for the vector of its means, and $\bB = \Cov(\bm X)$ for its
$k \times k$ covariance matrix.
Every linear statistic of the order statistics inherits its moments from $\balpha$ and
$\bB$. For $L = \sum_j w_j X_{(r_j)} = \bw'\bm X$,
\begin{equation}
\Ex[L] = \bw'\balpha, \qquad \Var(L) = \bw'\bB\bw .
\label{eq:linear}
\end{equation}
For several statistics with a weight matrix $\bm W$, one column per statistic, the
covariance matrix is $\bm W'\bB\bm W$. For example, a sample quantile under any of the nine
definitions of \citet{hyndman1996sample}, which are those of \proglang{R}'s
\code{quantile()}, is $w_1 X_{(j)} + w_2 X_{(j+1)}$ with fixed weights. So \eqref{eq:linear}
can be applied to it.

Another example in which the moment calculations are critical is Lloyd's BLUE. Consider a
location-scale distribution family with $X = \mu + \sigma Z$, where $Z$ is the standard
member of the family, with location parameter 0, scale parameter 1, and a known distribution,
and $\mu$ and $\sigma$ are unknown. We write $\balpha_Z$ and $\bB_Z$ for the mean vector and
the covariance matrix of $(Z_{(r_1)}, \dots, Z_{(r_k)})'$, the order statistics of $Z$ at
the same ranks, which the integrals above give for the distribution of $Z$. Then $\balpha =
\mu\bm 1 + \sigma\balpha_Z$ and $\bB = \sigma^2\bB_Z$, where $\bm 1$ is the vector of $k$
ones, and $\balpha_Z$ and $\bB_Z$ can be calculated independently of $\mu$ and $\sigma$. The
BLUE of \citet{lloyd1952least} from the ranks $r_1, \dots, r_k$ are the generalized
least-squares estimators (they exist when $\bB_Z$ is invertible and the matrix $\bm A$ below
has rank two)
\begin{equation}
(\hat\mu, \hat\sigma)' = (\bm A'\bB_Z^{-1}\bm A)^{-1}\bm A'\bB_Z^{-1}\bm X,
\qquad \bm A = (\bm 1, \balpha_Z),
\label{eq:blue} 
\end{equation}
and the covariance matrix of the two estimators is $\sigma^2(\bm A'\bB_Z^{-1}\bm A)^{-1}$. 

\subsection{Why direct numerical integration fails} \label{sec:background}

Applying \proglang{R}'s integrator directly to the integrals in \eqref{eq:moments} fails at
large $n$, and this subsection explains why. The integrand concentrates in a narrow region
that the integrator can miss, and the arithmetic of the computer overflows, underflows, or
loses accuracy by cancellation. Appendix~\ref{app:background} gives further background.

\paragraph{Properties of order statistics.} Two properties of the order statistics, their
center and their width, show where the integrands in \eqref{eq:moments}, $f_{r:n}$ and
$f_{r,s:n}$, have their mass. For that, we consider the probability transform
$U = F(X)$, which is uniform on $(0, 1)$ when $F$ is continuous. Note that
$X_{(r)} = Q(U_{(r)})$, where $U_{(r)}$ is the $r$th order statistic of the $n$ uniform
variables $F(X_i)$ and has the $\mathrm{Beta}(r, n-r+1)$ distribution
\citep[Chapter~2]{david2003order}. So $U_{(r)}$ has mean and standard deviation
\begin{equation}
p_r = \frac{r}{n+1}, \qquad s_r = \sqrt{\frac{p_r (1 - p_r)}{n + 2}},
\label{eq:beta}
\end{equation}
so $X_{(r)}$ concentrates near $c_r = Q(p_r)$, which we call its \emph{center}. Expanding
$Q$ about $p_r$ by the delta method, for large $n$ we have $X_{(r)} \approx c_r + (U_{(r)} -
p_r)/f(c_r)$, so the standard deviation of $X_{(r)}$ is approximately
\begin{equation}
\mathrm{SD}(X_{(r)}) \approx \frac{s_r}{f(c_r)},
\label{eq:width}
\end{equation}
which we call the \emph{width} of $X_{(r)}$, i.e., its large-sample standard deviation. 

How the width behaves as $n$ grows depends on the rank (Appendix~\ref{app:width}). For a
central rank, where $r/n \to p$ with $0 < p < 1$ and $0 < f(Q(p)) < \infty$, the width
shrinks like $n^{-1/2}$. For the minimum or the maximum, the approximation
\eqref{eq:width} is rougher, and the width depends on the tail of the parent distribution.
Similarly, the joint density $f_{r,s:n}$ concentrates near the point $(c_r, c_s)$, with the
widths of $X_{(r)}$ and $X_{(s)}$ in the two coordinates. The center and the width of every
integrand in \eqref{eq:moments} can be calculated from $Q$ and $f$ before any integration,
and our solutions in Section~\ref{sec:devices} rely on these two quantities.

\paragraph{\proglang{R}'s integrator and its blind spot.} We compute every integral with
\proglang{R}'s \code{integrate()} function, which applies the adaptive Gauss--Kronrod
routines of QUADPACK \citep{piessens1983quadpack, rcoreteam2023r}. The routine computes the
integrand at 21 fixed points of the range, called the nodes, and approximates the integral
by a weighted sum of these values, together with an estimate of the error. It then
repeatedly bisects the subinterval that has the largest error estimate and applies the same
rule to the two halves. Let $I$ denote the current value of the integral and $E$ denote the
sum of the error estimates. The routine stops when the number of subintervals reaches its
limit, or when
\begin{equation}
E \le \max(\mathrm{abs.tol},\ \mathrm{rel.tol} \times |I|),
\label{eq:stop}
\end{equation}
where \code{abs.tol} and \code{rel.tol} are arguments of \code{integrate()}, the absolute
and the relative tolerance, respectively. Appendix~\ref{app:integrate} describes the
procedure in more detail.

The Gauss--Kronrod rule has a blind spot in the calculation of the moments in
\eqref{eq:moments}, because it sees the integrand only at its nodes, which at first are
spread over the whole range. Our integrands, such
as $f_{r:n}$ at large $n$, are highly concentrated, and their mass can lie in a region
much narrower than the spacing of the nodes and so fall between them. In this case, the
rule returns a value and an error estimate that are both wrongly near zero, without
a warning message, because the stopping rule \eqref{eq:stop} is met at once. The Note of
the help page of \code{integrate()} warns of this: if the function is ``approximately
constant (in particular, zero) over nearly all its range'', then ``the result and error
estimate may be seriously wrong''.
For example, the density of the lower quartile of $n = \rpUnsplitN$ normal observations
integrates to one. When we apply \code{integrate()} to this density directly over the
whole real line, however, it returns $\rpUnsplitMass$, with a misleading error estimate of
$\rpUnsplitErr$ and the message \code{"OK"}.

\paragraph{Overflow, underflow and cancellation issues.} \proglang{R} stores a number as a
double-precision floating-point number (abbreviated to \emph{double} from here on). A double
carries about sixteen significant decimal digits, that is, the relative spacing of the
representable numbers is $\epsilon = 2^{-52} \approx 2.2 \times 10^{-16}$, and the largest
double is about $1.8 \times 10^{308}$ (Appendix~\ref{app:double}).

These limitations lead to three challenges. First, a product of many factors, such as the
coefficient and the powers in \eqref{eq:single}, can overflow or underflow even when the
product itself is of ordinary size. For example, the coefficient
$n!/\{(r-1)!\,(n-r)!\}$, when it is computed from its factorials, overflows from $n = 171$
on. Second, a difference of two nearly equal numbers keeps only the digits in which they
differ. This loss of accuracy of the difference is called \emph{cancellation}. Third, a
probability very close to one is stored with an absolute error of about $\epsilon$, so
$1 - F(x)$ formed from a stored $F(x)$ can have a very large relative error in the far upper
tail. For this reason \proglang{R}'s distribution functions return the upper tail
probability directly with \code{lower.tail = FALSE}, and its logarithm with
\code{log.p = TRUE}, which also keeps a tiny tail probability from underflowing. We call
$\log F(x)$ and $\log\{1 - F(x)\}$ the \emph{log-tails} and
$\log f(x)$ the \emph{log-density}.

\subsection{Our numerical design} \label{sec:devices}

This subsection explains the computing challenges and our solutions, in three groups: what
we compute, how we integrate, and how we evaluate the integrand. The first two groups are
our special design for the order statistics. The third is a critical but relatively
standard computing practice. Appendix~\ref{app:design} gives the further details of each
solution under the same headings, and Table~\ref{tab:challenges} there maps every challenge
to its solution.

\subsubsection{What we compute}

We compute the moments in standardized units, each about the center of its own order
statistic. The standardized units adapt the computation to the scale of the parent
distribution, and the centering avoids the cancellation in a variance.

\paragraph{Computing based on the standardized units.} For a parent distribution of $X$
in \eqref{eq:single} with the quantile function $Q$, we apply integration in the
standardized units
\begin{equation}
t = \frac{x - Q(1/2)}{\tilde\sigma}, \qquad
\tilde\sigma = \frac{Q(3/4) - Q(1/4)}{2\,\Phi^{-1}(3/4)},
\label{eq:units}
\end{equation}
where $\Phi^{-1}$ denotes the quantile function of the standard normal distribution
$N(0,1)$. We refer to $\tilde\sigma$ as the \emph{spread} of $X$ (or of the parent
distribution). It is the interquartile range of $X$ divided by the interquartile range of
the standard normal distribution, $2\Phi^{-1}(3/4) = 1.349$. The choice of the spread has
several benefits. First, it needs only $Q$, so it exists for any continuous parent
distribution, including a parent distribution without a variance, such as the Cauchy
distribution. Second, it is exactly 1 for the standard normal, so $t = x$ under standard
normality, and the constants that we calibrated there, such as those of the cut in
\eqref{eq:split}, carry over to every other parent distribution. More importantly, the unit
of $t$ is the spread of $X$, so the computation in $t$ automatically adapts to the scale of
$X$. For example, an interval defined in $t$ is in units of the spread of $X$, and an
absolute tolerance of $10^{-13}$ in $t$ is $10^{-13}$ spreads of $X$. This brings
significant convenience to our computing solutions to be discussed below.

Accordingly, we define $T = \{X - Q(1/2)\}/\tilde\sigma$. The distribution function, the
density, and the quantile function of $T$ are
\begin{equation*}
F_T(t) = F\{Q(1/2) + \tilde\sigma t\}, \qquad
f_T(t) = \tilde\sigma f\{Q(1/2) + \tilde\sigma t\}, \qquad
Q_T(u) = \frac{Q(u) - Q(1/2)}{\tilde\sigma}.
\end{equation*}
The order statistics transform in the same way,
$T_{(r)} = \{X_{(r)} - Q(1/2)\}/\tilde\sigma$, so the densities of $T_{(r)}$ and of
$(T_{(r)}, T_{(s)})$ are \eqref{eq:single} and \eqref{eq:double} with $F_T$ and $f_T$ in
place of $F$ and $f$, and we write them $f^T_{r:n}$ and $f^T_{r,s:n}$. The center and the
width of Section~\ref{sec:background} apply to $T_{(r)}$ with $Q_T$ and $f_T$ in place of
$Q$ and $f$. To keep them apart from those of $X_{(r)}$, we write the center of $T_{(r)}$
as
\begin{equation*}
\tilde c_r = Q_T(p_r) = \frac{c_r - Q(1/2)}{\tilde\sigma},
\end{equation*}
and its width is $s_r/f_T(\tilde c_r) = s_r/\{\tilde\sigma f(c_r)\}$, which equals the
width of $X_{(r)}$ in spreads.

\paragraph{Formulating moments.} Calculating a variance by
direct subtraction of the moments of $T_{(r)}$ can suffer from the cancellation issue. Each
moment, i.e., its integral, is obtained at a relative tolerance, say $\epsilon_q = 10^{-11}$
(see the discussion of the choice of the tolerances below), and the variance needs a relative
error of this order too to be accurate. For large $n$, $T_{(r)}$ concentrates at $\tilde c_r$
with $\tilde c_r^2 \gg \Var(T_{(r)})$. Then
\begin{equation*}
\Ex[T_{(r)}^2] = \Var(T_{(r)}) + (\Ex[T_{(r)}])^2 \approx \tilde c_r^2, \qquad
(\Ex[T_{(r)}])^2 \approx \tilde c_r^2,
\end{equation*}
so each term is about $\tilde c_r^2$ and carries an absolute error of about
$\epsilon_q \tilde c_r^2$. Their difference, the variance, then has a relative error of about
\begin{equation}
\epsilon_q\, \frac{\tilde c_r^2}{\Var(T_{(r)})}.
\label{eq:cancel}
\end{equation}
This ratio grows with $n$: at the rate $n$ at a central rank other than the median, and like
$n^2$ for the minimum or the maximum of a parent distribution whose density is positive and
finite at a finite endpoint of its support (Appendix~\ref{app:cancel}). For example, for the
minimum of a uniform sample and $\epsilon_q = 10^{-11}$, the relative error
\eqref{eq:cancel} reaches one at about $n = 6 \times 10^5$. The variance is then
\emph{lost}, in the sense that the returned value can be zero or even negative. A covariance
can be lost in the same way.

Our solution is to calculate the moments of the difference between $T_{(r)}$ and its center
$\tilde c_r$. Specifically, we compute
\begin{equation}
\begin{aligned}
m_r &= \Ex[T_{(r)} - \tilde c_r] = \int (t - \tilde c_r)\, f^T_{r:n}(t)\, dt, \\
v_r &= \Ex[(T_{(r)} - \tilde c_r)^2] = \int (t - \tilde c_r)^2 f^T_{r:n}(t)\, dt, \\
v_{rs} &= \Ex[(T_{(r)} - \tilde c_r)(T_{(s)} - \tilde c_s)]
       = \iint_{y < z} (y - \tilde c_r)(z - \tilde c_s)\, f^T_{r,s:n}(y, z)\, dy\, dz,
\end{aligned}
\label{eq:centered}
\end{equation}
over the support of $T$, and the variances and covariances are
\begin{equation}
\Var(T_{(r)}) = v_r - m_r^2, \qquad
\Cov(T_{(r)}, T_{(s)}) = v_{rs} - m_r m_s .
\label{eq:assemble}
\end{equation}
The subtraction in \eqref{eq:assemble} no longer suffers from cancellation, because $m_r$ is
at most of the order of the width of $T_{(r)}$, so $m_r^2$ is at most comparable to
$\Var(T_{(r)})$ and the subtraction loses little accuracy. We checked this over five
built-in families (the normal, uniform, exponential, logistic, and Gumbel distributions) at
$n = 10$, $10^3$, and $10^5$ and the five-number ranks (i.e., the ranks
of the minimum, the lower and upper quartiles, the median, and the maximum, given in
Section~\ref{sec:closedforms}): $m_r^2$ is at most \rpCenterSqRatioMax{} of $\Var(T_{(r)})$.

To return to the original units, reversing \eqref{eq:units} gives:
\begin{equation*}
\mu_{r:n} = Q(1/2) + \tilde\sigma (\tilde c_r + m_r), \qquad
\sigma_{r,s:n} = \tilde\sigma^2 (v_{rs} - m_r m_s),
\end{equation*}
where we set $v_{rr} = v_r$ for a variance.

\subsubsection{How we integrate}

We design the integration with a careful split of the range around the peak, with
tolerances scaled to the order statistic, with rules that check for a wrong result and
refuse it, and with a change of variable next to an endpoint of the support where the
density of the parent distribution is infinite.

\paragraph{Creating integration splits around the peaks.} By \eqref{eq:width}, the mass of
the density $f^T_{r:n}$ of $T_{(r)}$ lies within a few multiples of its width
$s_r/f_T(\tilde c_r)$ around $\tilde c_r$. The region of meaningful mass shrinks as $n$
grows, causing the blind spot of \code{integrate()} described in
Section~\ref{sec:background}.

Our solution is to cut the range according to the properties of $T_{(r)}$. A typical cut
splits the domain of $T_{(r)}$ into three \emph{pieces} at the two cut points
\begin{equation}
\tilde c_r \pm h_r, \qquad \text{ with } h_r = d_r + \frac{4}{\sqrt{n+1}}, \qquad
d_r = \begin{cases}
  10\, s_r / f_T(\tilde c_r) & \text{if } f_T(\tilde c_r) > 0, \\
  0 & \text{if } f_T(\tilde c_r) = 0,
\end{cases}
\label{eq:split}
\end{equation}
where $s_r$ is as in \eqref{eq:beta}, and $s_r/f_T(\tilde c_r)$ is the width of $T_{(r)}$.
We apply \code{integrate()} to the three pieces separately, summing their values and their
error estimates. We set $d_r$ to ten widths of $T_{(r)}$ for sufficient coverage, as the
densities of the order statistics at the smallest and the largest ranks are often skewed and
reach a few widths further on one side. The term $4/\sqrt{n+1}$ is a lower bound on $h_r$
and keeps the middle piece from collapsing where $f_T(\tilde c_r)$ is large ($d_r$ is then
small) or where $f_T(\tilde c_r)$ is zero. For the normal distribution, this lower bound is
about three widths of the sample median, whose width is $\sqrt{\pi/(2n)}$.

This cut can be used for calculating the second moment $v_r$ in \eqref{eq:centered}. The
peak of $f^T_{r:n}$ sits inside the finite middle piece, so the 21 nodes of
\code{integrate()} are allocated over a short interval around the peak, where the adaptive
bisection can find and resolve it. The two outer
pieces cover the lower tail and the upper tail.

The first moment $m_r$ in \eqref{eq:centered} needs one more cut. Its integrand
$(t - \tilde c_r) f^T_{r:n}(t)$ is negative for $t < \tilde c_r$ and positive for
$t > \tilde c_r$, and the sum of the two parts can suffer from the cancellation issue
because $m_r$ is small. Therefore, we cut the
range at $\tilde c_r$ as well, so $m_r$ runs on four pieces, each with an integrand of a
fixed sign.

The product moment $v_{rs}$ is a nested quadrature: an outer integral over $y$ whose
integrand is an inner integral over $z$, which is computed by one call of \code{integrate()}
at each outer node $y$. We cut the range of the inner integral at the cut points of rank
$s$, that is, $\tilde c_s \pm h_s$ and $\tilde c_s$. We cut the range of the outer integral
at the cut points of both ranks: $\tilde c_r \pm h_r$ and $\tilde c_r$, as well as $\tilde
c_s \pm h_s$ and $\tilde c_s$ to match the cuts of the inner integral. With this matching,
the inner integral keeps the same pieces while $y$ stays within one piece of the outer
range, so the computed outer integrand does not jump inside a piece
(Appendix~\ref{app:split}). Section~\ref{sec:identities} reports an example where one
product moment went wrong when the outer integral was cut at the cut points of rank $r$
alone.

\paragraph{Setting proper tolerances.} We set the two tolerances in \eqref{eq:stop} for each
integral, in the standardized units. The relative tolerance is $10^{-11}$, except for the
outer integral of a product moment, where it is $10^{-9}$ because the outer integrand
carries the error of the inner integral as noise. The absolute tolerances lie between
$10^{-15}$ and $10^{-12}$. However, a fixed absolute tolerance can exceed the moment that it
is meant to compute. For example, the variance of the minimum of $10^6$ uniform observations
is $7 \times 10^{-12}$ in standardized units. We therefore multiply each absolute tolerance
by a factor that is smaller than one when the width of the order statistic is small, so that
the tolerance tightens for such an order statistic. We also tighten the inner tolerance of a
product moment in the far tail of a heavy-tailed parent distribution, where the outer factor
$y - \tilde c_r$ magnifies the inner error. Appendix~\ref{app:tol} gives the values and the
reasons for them.

\paragraph{Setting proper acceptance and refusing rules.} As Section~\ref{sec:background}
shows, \code{integrate()} may return a wrong value with a small error estimate and no
warning message. Conversely, a warning message may not indicate a wrong value. Therefore, we
accept or refuse a result based on five rules, not just on the messages
(Appendix~\ref{app:rules}). Two of the rules test the result itself. The first is the
\emph{mass check}. For every order statistic whose mean is computed by quadrature, we also
integrate its density, with the same cuts, and we require this integral, the total mass, to
equal one within $10^{-7}$. A mass short of one shows that a part of the density was missed,
for example a second mode of the parent distribution that falls between the nodes. The
second is the \emph{variance guard}. It stops the computation when a variance comes out zero
or negative, which is how a variance lost by the cancellation in \eqref{eq:cancel} shows. The other three rules decide what to do with a non-finite value of the
functions of the parent distribution and with a warning message of \code{integrate()}. In
particular, they refuse a moment that does not exist, which \code{integrate()} may return as
a finite value.

\paragraph{Integrating next to an endpoint where the density is infinite.} For some
parent distributions, such as a gamma or a Weibull distribution with a shape parameter
below one, the density is infinite at a finite endpoint of the support. Write $\delta$ for
the distance from such an endpoint, in the standardized units of $t$. The density then
behaves like $\delta^{\theta - 1}$ with an index $0 < \theta < 1$, which is the shape
parameter in these two examples. This causes two difficulties in $t$. First,
\code{integrate()} does not resolve such a singularity to our tolerances. Second, the
origin of $t$ is the median, so the endpoint is a number of ordinary size, and an order
statistic at an extreme rank can lie closer to the endpoint than the spacing of the doubles
there. Our solution is to integrate the pieces next to such an endpoint in the variable
$\eta = \delta^{\theta}$. In this variable the integrand is bounded. Moreover, the distance
$\delta = \eta^{1/\theta}$ is computed from the integration variable and not from $t$, so a
tiny distance keeps its relative precision. Appendix~\ref{app:endpoint} gives the details,
and Section~\ref{sec:newfamilies} gives the evidence.

\subsubsection{How we evaluate the integrand}

We compute the integrand on the logarithmic scale, and we form the difference
$F(z) - F(y)$ with care to keep its precision when $y$ and $z$ lie in the same tail.

\paragraph{Computing at the logarithmic scale.} The coefficient and the powers in
\eqref{eq:single} and \eqref{eq:double} overflow or underflow at large $n$
(Section~\ref{sec:background}), although their product is of ordinary size. We therefore
compute the logarithm of the integrand and exponentiate it afterwards:
\begin{equation}
\log f^T_{r:n}(t) = -\log \mathrm{B}(r, n-r+1) + (r - 1)\log F_T(t)
+ (n - r)\log\{1 - F_T(t)\} + \log f_T(t),
\label{eq:logint}
\end{equation}
where $\mathrm{B}(a, b) = \Gamma(a)\Gamma(b)/\Gamma(a+b)$ is the beta function, so that
$1/\mathrm{B}(r, n-r+1)$ is the coefficient. The joint density is treated in the same way.
The log-tails and the log-density come directly from the distribution function and the
density function of the parent distribution, through their arguments \code{log.p},
\code{lower.tail}, and \code{log}. Computing on the logarithmic scale follows the approach
of \citet{royston1982algorithm} for the means of normal order statistics. We apply the same
idea to the product moments and to any parent distribution whose functions honor
\code{log.p} and \code{lower.tail} (Appendix~\ref{app:log}).

\paragraph{Computing the difference $F(z) - F(y)$.} The factor $\{F(z) - F(y)\}^{s-r-1}$ in
\eqref{eq:double} needs care when $y$ and $z$ lie in the same tail. In the far upper tail,
for example, $F(y)$ and $F(z)$ are both near one, and their difference loses its accuracy by
cancellation. We therefore form the difference on the side where the probabilities are
small. When both points lie below the median of the parent distribution, we compute $F(z)\{1
- F(y)/F(z)\}$. When both points lie above the median of the parent distribution, we compute
$S(y)\{1 - S(z)/S(y)\}$, where $S = 1 - F$ denotes the upper tail probability. Each ratio is
obtained from the log-tails. When the two points lie on different sides of the median, we
compute $F(z) - F(y)$ directly. Appendix~\ref{app:diff} gives the error analysis.

\Needspace*{4\baselineskip}
\subsection{Closed forms, the moment condition and the internal checks} \label{sec:closed}

Four of the built-in families have closed forms for some moments of their order
statistics. For these moments, the package returns the values of the closed forms in place
of the quadrature, and we also use the forms to check the accuracy of the quadrature
(Section~\ref{sec:closedforms}). For $r \le s$, the closed forms of the moments for the
uniform distribution on $(0, 1)$, the exponential distribution with unit rate, the standard
logistic distribution, and the Gumbel distribution (with $F(x) = \exp(-e^{-x})$) are,
respectively, 
\begin{align*}
\text{uniform:} \quad & \mu_{r:n} = \frac{r}{n+1}, &
  \sigma_{r,s:n} &= \frac{r\,(n-s+1)}{(n+1)^2 (n+2)}, \\
\text{exponential:} \quad & \mu_{r:n} = \sum_{i=1}^{r} \frac{1}{n-i+1}, &
  \sigma_{r,s:n} &= \sigma_{r,r:n} = \sum_{i=1}^{r} \frac{1}{(n-i+1)^2}, \\
\text{logistic:} \quad & \mu_{r:n} = \psi(r) - \psi(n-r+1), &
  \sigma_{r,r:n} &= \psi'(r) + \psi'(n-r+1), \\
\text{Gumbel:} \quad & \mu_{n:n} = \gamma + \log n, &
  \sigma_{n,n:n} &= \frac{\pi^2}{6},
\end{align*}
where $\psi = \Gamma'/\Gamma$ denotes the digamma function, $\psi'$ denotes its derivative,
and $\gamma$ represents Euler's constant. Appendix~\ref{app:closed} gives the sources of
these forms and the facts behind them. The normal distribution has no closed form at
general $n$ and is computed by quadrature throughout.

A moment is computed only if it exists, and the condition depends on the rank. Consider a
parent distribution whose density, far out in each tail, lies between two constant multiples
of $|x|^{-\nu_L - 1}$ in the lower tail and of $x^{-\nu_U - 1}$ in the upper tail, for some
$\nu_L, \nu_U > 0$. We call $\nu_L$ and $\nu_U$ its \emph{tail indices}. For example,
Student's $t$ distribution with $\nu$ degrees of freedom has $\nu_L = \nu_U = \nu$. The
density \eqref{eq:single} of $X_{(r)}$ then decays like $|x|^{-\nu_L r - 1}$ in the lower
tail and like $x^{-\nu_U (n-r+1) - 1}$ in the upper tail. The reason is that, in the lower
tail, $F(x)$ decays like $|x|^{-\nu_L}$, so $F(x)^{r-1} f(x)$ decays like
$|x|^{-\nu_L (r-1) - \nu_L - 1}$, while $\{1 - F(x)\}^{n-r}$ tends to one. The upper tail
is the same with $1 - F$ in place of $F$ (Appendix~\ref{app:closed} gives the derivation).
So, for a moment of order $k > 0$,
\begin{equation*}
\Ex|X_{(r)}|^k < \infty \quad \text{if and only if} \quad
\nu_L r > k \ \text{ and } \ \nu_U (n - r + 1) > k .
\end{equation*}
The normal distribution does not have this form, because its density decays faster than any
power of $|x|$ in both tails. A tail of this kind, or a bounded tail such as the lower tail
of the exponential distribution, is given the tail index $\infty$, and the condition then
holds in that tail for every $k$ and every rank. Among the other built-in families
(Table~\ref{tab:families}), the Cauchy distribution has $\nu_L = \nu_U = 1$, and the $F$
distribution with $\nu_1$ and $\nu_2$ degrees of freedom has $\nu_L = \infty$ and
$\nu_U = \nu_2/2$. The tail indices of the remaining ten families are all $\infty$.

The package returns the variance of every requested rank, so it applies this condition
with $k = 2$ to every requested rank, and it
refuses a rank whose second moment does not exist. The covariances then exist as well, by
the Cauchy--Schwarz inequality. For a built-in family the package sets the tail indices
itself. For a user-supplied distribution it uses the tail indices that the user declares,
and it does not check them (Section~\ref{sec:summary}). For example, the median of five
Cauchy observations has a variance, because
$\nu_L r = \nu_U (n - r + 1) = 3 > 2$ at $r = 3$. In contrast, the minimum and the maximum
of a Cauchy sample have no mean at any sample size.

The package internally checks its own results in two ways. The first is reflection symmetry.
Suppose that the parent distribution is symmetric about its median and that the rank set
contains the rank $n + 1 - r$ whenever it contains the rank $r$, as the complete rank set
$1, \dots, n$ does. Then the means of the order statistics at the ranks $r$ and $n + 1 - r$
lie symmetrically about the median of the parent distribution, and their second moments
about the median are equal. For example, for the standard normal distribution at $n = 5$,
$\mu_{1:5} = -\mu_{5:5}$ and $\Ex[X_{(1)}^2] = \Ex[X_{(5)}^2]$. The integrators do
not impose this symmetry, so a violation would signal a quadrature failure. The second is
the permutation identities. For the complete rank set $r = 1, \dots, n$ and a parent
distribution with a finite second moment, the order statistics are a permutation of the
sample, so
\begin{equation}
\sum_r \mu_{r:n} = n\,\Ex[X], \qquad
\sum_r \Ex[X_{(r)}^2] = n\,\Ex[X^2], \qquad
\sum_{r,s} \sigma_{r,s:n} = n\,\Var(X).
\label{eq:identities}
\end{equation}
For the normal distribution, every row of $\bB$ also sums to the variance of the parent
distribution, at every $n$, because the sample mean $\bar X$ is independent of the
deviations $X_{(r)} - \bar X$ (Appendix~\ref{app:closed}). This row-sum identity does not
hold for other parent distributions in general, so the package checks it for the normal
distribution only. The package stores the residuals of these checks with the result, that
is, the absolute differences between the two sides of each identity, in the standardized
units in \eqref{eq:units}. A residual above $10^{-6}$ raises a warning. When a calculated
covariance matrix has a negative eigenvalue beyond rounding, a warning is also given.

The package also reports the sum of the error estimates that \code{integrate()} returned
for the single integrals and the outer integrals. That sum is just an estimate of the total
error (Appendix~\ref{app:closed}). More reliable evidence of accuracy lies in the agreement
with the closed forms, the identities in \eqref{eq:identities}, and the printed tables in
the literature (Section~\ref{sec:accuracy}).

\Needspace*{4\baselineskip}
\section[Using the orderStatMoments package]{Using the \pkg{orderStatMoments} package} \label{sec:package}

This section shows, from the user's side, how to compute the moments and the statistics
built on them, with two illustrations. Appendix~\ref{app:examples} gives four further
examples. The package only depends on base \proglang{R} and \pkg{stats}
\citep{rcoreteam2023r}. It is installed from its repository with
\begin{Code}
R> remotes::install_github("ZWuLab/orderStatMoments", build_vignettes = TRUE)
\end{Code}

\Needspace*{4\baselineskip}
\subsection{The main call} \label{sec:call}

A call to \code{os_moments()} requires the sample size, the needed ranks (every rank by
default), and the distribution with its parameters. The constants \code{os_shift} $= a$ and
\code{os_multiplier} $= b$ of $Y = a + bX$ in Section~\ref{sec:integrals} are optional. The result is an object of class
\code{"orderStatMoments"}. Its print method shows the means, the covariance matrix, whether
the numbers come from closed forms or from quadrature, the summed error estimate, and the
run-time checks.
\needspace{4\baselineskip}   %
\begin{CodeChunk}
\begin{CodeInput}
R> library("orderStatMoments")
R> m <- os_moments(n = 20, ranks = c(1, 5, 10, 20), dist = "norm")
R> m
\end{CodeInput}
\begin{CodeOutput}
Exact moments of order statistics: norm (mean = 0, sd = 1), n = 20,
  ranks 1 5 10 20
alpha (means):
         r1          r5         r10         r20 
-1.86747500 -0.74538300 -0.06199629  1.86747500 
B (covariance matrix):
            r1         r5        r10        r20
r1  0.27569660 0.05705664 0.02965625 0.01020472
r5  0.05705664 0.09399600 0.04933743 0.01707114
r10 0.02965625 0.04933743 0.07694744 0.02688388
r20 0.01020472 0.01707114 0.02688388 0.27569660
route: integrate; quadrature error estimate: 1.5e-11
checks: reflection n/a; identities n/a
\end{CodeOutput}
\end{CodeChunk}
 An object of class \code{"orderStatMoments"} includes the means named by rank in
\code{alpha}, the variances in \code{var}, and the covariance matrix in \code{B}, with the
variances on its diagonal. The print method does not show \code{var} beside \code{B},
because the variances are the diagonal of \code{B}, and it shows \code{var} in place of
\code{B} in a call with \code{cov = FALSE}. Four more fields describe the computation.
\code{route} says whether the numbers come from closed forms, from quadrature, or from both.
\code{quad_err} is the summed error estimate of Section~\ref{sec:closed}. \code{checks}
holds the residuals of the run-time checks. And \code{standard} holds the moments of the
family's standard member, which Section~\ref{sec:families} defines. Here, both checks print as not applicable,
because the rank set $\{1, 5, 10, 20\}$ at $n = 20$ is not complete, and it does not contain
$n + 1 - r$ for every rank $r$ in it (it contains 5 but not 16). When only the means and the
variances are wanted at many ranks, users should set \code{cov = FALSE} to skip the double
integrals for more efficient computing.

When every rank is requested, which is the default, the result is the full $n \times n$
covariance matrix, and the run-time checks of Section~\ref{sec:closed} apply to it
(Appendix~\ref{app:full} shows an example).

\Needspace*{4\baselineskip}
\subsection{The families and the closed forms} \label{sec:families}

Thirteen families are built in: the twelve continuous distributions for which base
\proglang{R} provides the distribution, density, and quantile functions, and the
Gumbel distribution. A family is specified as in
\proglang{R}. The argument \code{dist} is the name that \proglang{R}'s functions carry after
\code{p}, \code{d}, and \code{q} (\code{"gamma"} for \code{pgamma()}), and the parameters
are passed under \proglang{R}'s own names, with \proglang{R}'s defaults
(Table~\ref{tab:families}). A user who writes \code{rgamma(n, shape = 2, rate = 3)} to
draw a sample writes \code{os_moments(n, dist = "gamma", shape = 2, rate = 3)} for the
moments of its order statistics.

\begin{table}[t!]
\centering
\small
\begin{tabular}{l >{\raggedright\arraybackslash}p{5.2cm} >{\raggedright\arraybackslash}p{6.4cm}}
\toprule
\code{dist} & Parameters, with their defaults & Closed forms returned by default \\
\midrule
\code{"norm"} & \code{mean = 0}, \code{sd = 1} & none; every moment by quadrature \\
\code{"unif"} & \code{min = 0}, \code{max = 1} & every mean, variance, and covariance \\
\code{"exp"} & \code{rate = 1} & every mean, variance, and covariance \\
\code{"logis"} & \code{location = 0}, \code{scale = 1} & the means and the variances \\
\code{"gumbel"} & \code{location = 0}, \code{scale = 1} & the mean and variance of the sample maximum $X_{(n)}$ \\
\midrule
\code{"t"} & \code{df} & none \\
\code{"cauchy"} & \code{location = 0}, \code{scale = 1} & none \\
\code{"gamma"} & \code{shape}, and \code{rate = 1} or \code{scale} & none \\
\code{"chisq"} & \code{df} & none \\
\code{"beta"} & \code{shape1}, \code{shape2} & none \\
\code{"lnorm"} & \code{meanlog = 0}, \code{sdlog = 1} & none \\
\code{"weibull"} & \code{shape}, \code{scale = 1} & none \\
\code{"f"} & \code{df1}, \code{df2} & none \\
\bottomrule
\end{tabular}
\caption{\label{tab:families} The thirteen built-in families, their parameters under the
names and with the defaults of \proglang{R}, and the closed forms returned in place of the
quadrature. A parameter without a default must be given. The Gumbel distribution is in the
maximum form, $F(x) = \exp\{-e^{-(x - \mathrm{location})/\mathrm{scale}}\}$.}
\end{table}

The quadrature depends only on the shape of a distribution, so the parameters of a
family are treated in two groups. The shape parameters (\code{df},
\code{shape}, \code{shape1}, \code{shape2}, \code{df1}, \code{df2}, and \code{sdlog}) select
the family's \emph{standard member}, which is the member with the shape parameters of the
call and \proglang{R}'s defaults for the other parameters. The package computes the moments
of the standard member. The other parameters determine a shift and a multiplier, which the
package applies exactly afterwards. For example, the mean and
the standard deviation of the normal distribution are a shift and a multiplier, and the rate
of the gamma distribution is the reciprocal of a multiplier. The user therefore does not
need to know which parameter does what, and a large mean or a small rate does not enter the
quadrature. The arguments \code{os_shift} $= a$ and \code{os_multiplier} $= b$ serve a
family that has no parameter of its own for a shift or a multiplier, such as Student's $t$
distribution, which has only \code{df}. They give the moments of $a + bX$, where $X$ has
the distribution that the family's own parameters specify
(Appendix~\ref{app:bounded} shows two examples). The noncentral distributions
(\proglang{R}'s \code{ncp}) are not built in.

The package also sets the support and the tail indices of each family
(Section~\ref{sec:closed}), so a moment that does not exist is refused without a
declaration by the user. The Cauchy distribution shows this moment condition at work. The
median of five Cauchy observations has a variance, but the order statistics at the ranks 1,
2, 4, and 5 have no second moment, so the package stops at the first such rank:
\needspace{8\baselineskip}
\begin{CodeChunk}
\begin{CodeInput}
R> os_moments(5, ranks = 3, dist = "cauchy")$var
R> os_moments(5, dist = "cauchy")
\end{CodeInput}
\begin{CodeOutput}
      r3 
1.221253 
Error: os_moments(): the moment of order 2 of X_(1) does not exist:
  the lower tail index 1 times the rank 1 is not above 2
\end{CodeOutput}
\end{CodeChunk}
 
By default the package uses a closed form wherever the family has one, and the
\code{route} field reports it (Appendix~\ref{app:exp} shows an example). The argument
\code{method = "integrate"} forces the quadrature everywhere, and the study of
Section~\ref{sec:closedforms} uses it to check the quadrature accuracy against the closed
forms.
With \code{method = "closed"}, the package refuses a quantity that has no closed form.

\subsection{A user-supplied distribution} \label{sec:userdist}

A distribution that is not built in enters through \code{os_dist()}, from its
distribution, density, and quantile functions in \proglang{R}'s convention. For example,
the Pareto distribution with shape three on $[1, \infty)$, whose survival function is
$x^{-3}$, is built like this:
\needspace{10\baselineskip}
\begin{CodeChunk}
\begin{CodeInput}
R> pareto3 <- os_dist(
+    p = function(q, lower.tail = TRUE, log.p = FALSE) {
+      ls <- -3 * log(pmax(q, 1))
+      if (lower.tail) { if (log.p) log(-expm1(ls)) else -expm1(ls) }
+      else if (log.p) ls else exp(ls)
+    },
+    d = function(x, log = FALSE) {
+      ld <- ifelse(x >= 1, log(3) - 4 * log(x), -Inf)
+      if (log) ld else exp(ld)
+    },
+    q = function(p) (1 - p)^(-1/3),
+    support = c(1, Inf), tail = c(Inf, 3), name = "pareto3")
R> os_moments(5, dist = pareto3, cov = FALSE)
\end{CodeInput}
\begin{CodeOutput}
Exact moments of order statistics: pareto3, n = 5, ranks 1 2 3 4 5
alpha (means):
      r1       r2       r3       r4       r5 
1.071429 1.168831 1.314935 1.577922 2.366883 
var (variances; cov = FALSE):
        r1         r2         r3         r4         r5 
0.00588697 0.01844908 0.05116556 0.18049160 2.40885300 
route: integrate; quadrature error estimate: 3.14e-12
checks: reflection n/a; identities sum_alpha 0.00e+00, sum_sq 7.11e-15
\end{CodeOutput}
\end{CodeChunk}
 The distribution function must honor \code{log.p} and \code{lower.tail}, and the density
must honor \code{log}, because the integrands are evaluated on the log scale in both tails.
\code{os_dist()} checks at a few points that the three functions agree with each other. The
argument \code{support} gives the endpoints of the support, which are the limits of
integration. The tail indices $(\nu_L, \nu_U)$ of Section~\ref{sec:closed} decide which
moments exist. The user declares them in the argument \code{tail}, here
\code{tail = c(Inf, 3)}, because the lower tail is bounded and the density decays like
$x^{-4} = x^{-\nu_U - 1}$ with $\nu_U = 3$ in the upper tail. Its default, \code{c(Inf, Inf)},
is for a distribution whose tails are bounded or decay faster than any power. With the
default left in place for a heavy-tailed distribution, the package cannot be relied on to
refuse a moment that does not exist.

A user-supplied distribution has no parameters to pass. Its three functions fix the
distribution of $X$ in Section~\ref{sec:integrals}, as the functions of \code{pareto3} above
fix the shape at three, and the arguments \code{os_shift} and \code{os_multiplier} give the
moments of $a + bX$. For the BLUE of Section~\ref{sec:linear}, the supplied distribution is
the standard member $Z$. Appendix~\ref{app:laplace} gives a second example, the Laplace
distribution. It shows the two other arguments of \code{os_dist()}: \code{symmetric}, for a
symmetric distribution, and \code{breaks}, for the points inside the support where the
density is not smooth.

\subsection{Linear statistics of order statistics, with two illustrations} \label{sec:linear}

Three functions turn the moments of order statistics into the moments of the statistics
a user computes from them, by \eqref{eq:linear} and \eqref{eq:blue}.
\code{os_quantile_weights(p, n, type)} returns the ranks and the weights of the sample
quantiles at the probabilities \code{p} under any of the nine definitions of
\code{quantile()}, one column of weights per probability.
\code{os_lin_moments(w, mom)} takes a weight matrix with one column per statistic, and an
\code{orderStatMoments} object at the ranks that the rows name. It returns the means, the
variances, and the covariance matrix of the statistics. \code{os_blue(mom)} returns
Lloyd's weights for the location parameter $\mu$ and the scale parameter $\sigma$ in
\eqref{eq:blue}, \code{w_mu} and \code{w_sigma}. It also returns the variances and the
covariance of the two estimators per unit of $\sigma^2$, \code{var_mu}, \code{var_sigma},
and \code{cov}, so that, for example, the variance of $\hat\mu$ is $\sigma^2$ times
\code{var_mu}.
\code{os_blue()} works with the moments of the standard member $Z$, which it reads from the
field \code{standard}, so the shift and the multiplier in
the call of \code{os_moments()} do not affect its result.

We illustrate these functions on two cases that a user meets: the mean and the standard
deviation estimated from a five-number summary, and the median of an even sample. In each
case, we compute the weights of an estimator and its sampling variance from the moments of
the order statistics. Appendix~\ref{app:logistic} adds a third case, the BLUE for a
logistic sample.

\subsubsection{A five-number summary of a normal sample} \label{sec:five}

Suppose a study reports only the minimum, the three quartiles, and the maximum of
$n = \rpFiveN$ observations, with the quartiles under \proglang{R}'s default definition
(type 7). The five values are then the order statistics at the ranks
$1, 7, 13, 19, 25$. If the data are normal, the BLUE of the population mean $\mu$ and the
population standard deviation $\sigma$ from these five values are given by
\eqref{eq:blue} at those ranks:
\needspace{10\baselineskip}
\begin{CodeChunk}
\begin{CodeInput}
R> five <- c(1, 7, 13, 19, 25)
R> b <- os_blue(os_moments(25, ranks = five))
R> round(b$w_mu, 4)
R> round(b$w_sigma, 4)
R> c(sd_of_mean = sqrt(b$var_mu), sd_of_sd = sqrt(b$var_sigma))
\end{CodeInput}
\begin{CodeOutput}
    r1     r7    r13    r19    r25 
0.0821 0.2857 0.2644 0.2857 0.0821 
     r1      r7     r13     r19     r25 
-0.1713 -0.2566  0.0000  0.2566  0.1713 
sd_of_mean   sd_of_sd 
 0.2068675  0.1585884 
\end{CodeOutput}
\end{CodeChunk}
 The location estimator puts weight \rpFiveWLocMedian{} on the median and
\rpFiveWLocExtreme{} on each of the minimum and the maximum, with the remaining weight on
the two quartiles. The weights of the scale estimator are antisymmetric, as they must be for
a parent distribution that is symmetric about zero and a rank set that contains $n + 1 - r$
with every rank $r$. The standard error of the estimated mean is
$\rpFiveSeMeanFactor\,\sigma$, against $\sigma/\sqrt{n} = \rpFiveSeSampleMean\,\sigma$ for
the sample mean of all \rpFiveN{} observations. So the standard error from the five-number
summary is \rpFiveSeMeanPct\% larger. The standard error of the estimated standard
deviation is $\rpFiveSeSdFactor\,\sigma$. 

\subsubsection{The median of an even sample} \label{sec:median} 

How do the medians of ten and of eleven normal observations compare in variance? Under
\proglang{R}'s default definition, the median of ten is the average of the fifth and
sixth order statistics, and its moments follow from theirs by \eqref{eq:linear}:
\needspace{4\baselineskip}   %
\begin{CodeChunk}
\begin{CodeInput}
R> qw <- os_quantile_weights(p = 0.5, n = 10)
R> qw$ranks
R> qw$W
R> med10 <- os_lin_moments(qw$W, os_moments(10, ranks = qw$ranks))
R> m11 <- os_moments(11, ranks = 6)
R> c(var_even_10 = med10$var[[1]], var_odd_11 = m11$var[[1]])
\end{CodeInput}
\begin{CodeOutput}
[1] 5 6
   50%
r5 0.5
r6 0.5
var_even_10  var_odd_11 
  0.1383264   0.1371624 
\end{CodeOutput}
\end{CodeChunk}
 The weights are \rpMedTenWeight{} on the ranks \rpMedTenRanks. For the standard normal
distribution,
the variance of the median of ten is \rpMedTenVar, against \rpMedElevenVar{} for the
median of eleven, which is a single order statistic. So the median of ten has
\rpMedRatio{} times the variance of the median of eleven. The same two functions give
the moments of any sample quantile under any of the nine definitions, and of any linear
combination of order statistics, such as a trimmed mean or an interquartile range,
exactly at the sample size at hand.

\section{Accuracy} \label{sec:accuracy}

This section gives the evidence that the quadrature returns the exact moments. The evidence
comes in three layers that rest on different grounds: (1) the closed forms of four families
are exact values computed apart from the quadrature; (2) the permutation identities on full
matrices hold for every parent distribution with a finite second moment; and (3) the printed tables are independent
computations in the literature. Every comparison in this section ran the package with \code{method =
"integrate"}, so that each number tests the quadrature and not a closed form. A residual is
the absolute difference between two values that should be equal, such as the quadrature and
a closed form, or the two sides of an identity. Unless stated otherwise, a residual or a
difference is in the units of the family's standard member. Appendix~\ref{app:accuracy}
gives further results for each of the three layers. The three layers concern five of the
built-in families, and Section~\ref{sec:newfamilies} gives the evidence for the other eight.

\subsection{Closed forms} \label{sec:closedforms}

\begin{figure}[t!]
\centering
\includegraphics[width=\textwidth]{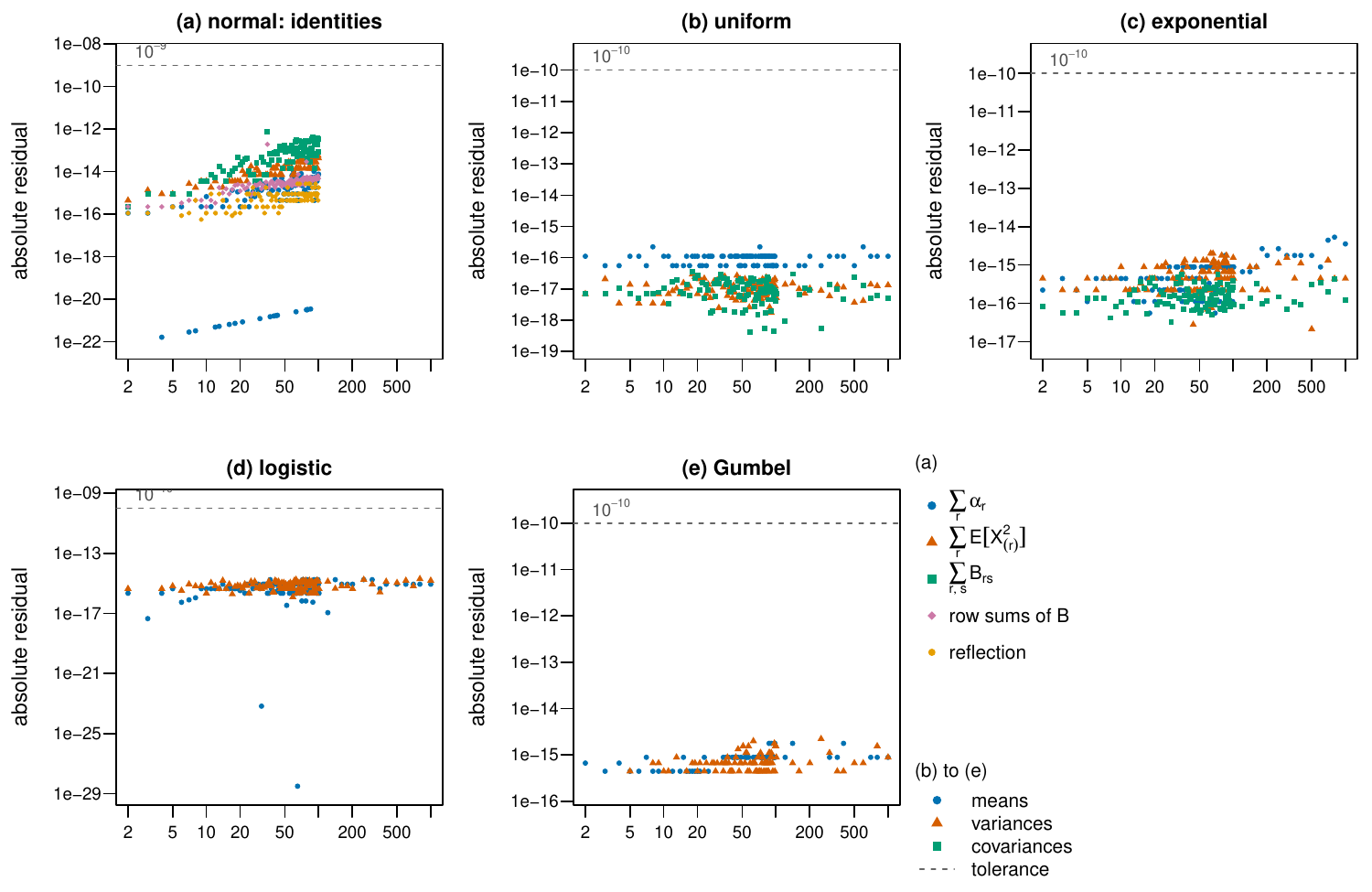}
\caption{\label{fig:resid} The residuals against $n$ on logarithmic axes, in the units of
the standard member, with the tolerance of each panel as a dashed line. Each point is the
largest residual at one $n$ for one of the quantities that the legend names. Panel (a): the
residuals of the checks on the full matrices of the normal distribution, at every $n$ from 2
to 100 (Section~\ref{sec:identities}). They are the absolute differences between the two
sides of each identity in \eqref{eq:identities} and between each row sum of $\bB$ and one.
For the reflection symmetry of Section~\ref{sec:closed}, they measure how far the means and
the second moments about the median at the ranks $r$ and $n + 1 - r$ depart from that
symmetry. Panels (b) to (e): the absolute differences between the quadrature and the closed
forms of the four families, over the means, the variances, and the covariances
(Section~\ref{sec:closedforms}). The \atFigZeroCount{}
residuals that are exactly zero have no logarithm and are not drawn.}
\end{figure}

The quadrature reproduces every closed form of Section~\ref{sec:closed} within
$10^{-10}$ at every one of \atClosedSizes{} sample sizes from 2 to \atClosedSizeMax,
including every integer from 2 to 100 and \atClosedSizesAboveHundred{} further sample sizes
up to \atClosedSizeMax. At each sample size, we check the five-number ranks, $1$,
$\lfloor n/4 \rfloor + 1$, $\lfloor n/2 \rfloor + 1$, $n - \lfloor n/4 \rfloor$, and $n$,
and all their pairs. At $n = 4m + 1$ for an integer $m$, these five ranks are those of the
minimum, the lower quartile, the median, the upper quartile, and the maximum.
Figure~\ref{fig:resid} shows every residual. The largest residual is \atClosedWorst, for
the \atClosedWorstQuantity{} of the \atClosedWorstFamily{} family at
$n = \atClosedWorstN$. This largest residual is a factor \atClosedMargin{} below $10^{-10}$.
Appendix~\ref{app:closedforms} gives the largest residual of each family
(Table~\ref{tab:closed}), and a check at $n = 2$ of the covariances of the logistic and
the Gumbel distributions, which have no closed form.

\paragraph{Beyond $n = \atClosedSizeMax$.} Where a closed form exists, the comparison can go
far past $n = \atClosedSizeMax$ and show how many digits survive at $n$ in the millions. We
repeat the comparison for the uniform and the exponential distributions at \atLargeNEq, at
the same five ranks (Appendix~\ref{app:closedforms}, Table~\ref{tab:largen}). There the
residuals are relative, because the variance of the minimum is of order $n^{-2}$ for both
distributions. The covariances are measured through the correlations $\rho_{rs} =
\sigma_{r,s:n}/(\sigma_{r,r:n}\sigma_{s,s:n})^{1/2}$. At every sample size, the means and
the variances are reproduced within \atLargeNMeanWorst{} and \atLargeNVarWorst{} in relative
terms, and the correlations within \atLargeNCorrWorst. A pair of order statistics that is
nearly independent keeps fewer digits of its covariance (Appendix~\ref{app:closedforms}
gives the reason). The minimum and the maximum, whose correlation is of order $1/n$, keep
\atLargeNExtDigits{} significant digits of their covariance, and every other pair keeps at
least \atLargeNOtherDigits significant digits.

\Needspace*{4\baselineskip}
\subsection{Identities on full matrices} \label{sec:identities}

The permutation identities \eqref{eq:identities} hold on every full matrix within $10^{-9}$.
The matrices are those of the normal distribution at every $n$ from 2 to 100, and those of
the uniform, exponential, logistic, and Gumbel families at $n = \atIdentOtherSizes$. Panel (a) of Figure~\ref{fig:resid}
shows the residuals of the normal distribution against $n$, and
Appendix~\ref{app:identities} gives the largest residual of each identity by family
(Table~\ref{tab:ident}). The largest residual is \atIdentWorstAll{} (\atIdentWorstAllFamily,
$n = \atIdentWorstAllN$). For the normal distribution, the unit row sums of $\bB$ hold
within \atIdentRowSumsWorst. The row sums are the sharpest of the identities, since each of
the $n$ rows is its own check.

The identities test sums, so they would pass errors that cancel in a sum. That is why they
are the second layer of evidence and not the first. Still, they caught the one failure found
in the runs of this section. In an earlier version of the package, the outer integral of a
product moment was cut only at the cut points of rank $r$. The product moment of the ranks
$(9, 20)$ at $n = 70$ was then off by several hundred times its own error estimate, and the
sum of its row of $\bB$ showed the error (Section~\ref{sec:devices} gives the fix).

For the normal distribution, the identities also carry the evidence at the top of the
range, where no printed table reaches. On the full covariance matrix at $n = \atLargeN$,
every one of the \atLargeN{} row sums lies within \atLargeRowMax{} of one, and the sum of
all entries is within \atLargeTotal{} of $n$ (Appendix~\ref{app:identities}).

\subsection{The printed tables in the literature} \label{sec:tables}

The package agrees with \cpCellsAgree{} of the \cpCellsTotal{} printed cells of
\countword{\cpPublicationsTotal} publications, where a cell is one number of a printed table
(Table~\ref{tab:tables}). On each of the
\cpCellsTable{} cells that disagree, an independent check sides with the package, and we
therefore attribute each disagreement to the table. This check, which we call the
conditional-beta check, computes the same moment by integrating in the probability
variable $u = F(x)$ against beta densities, so it does not use the package's integrands
(Appendix~\ref{app:tables}). The criterion of agreement is the accuracy that each source
states. It is $10^{-7}$ for \citet{tietjen1976variances}, whose table prints ten decimals
and certifies seven. For the other sources it is one unit of the last printed decimal, or
one unit of the last significant figure where a table prints significant figures.

\begin{table}[t!]
\centering
\footnotesize
\setlength{\tabcolsep}{3.5pt}
\begin{tabular}{p{3.9cm}lp{2.7cm}lrrr}
\toprule
source & family & quantities & sizes $n$ & cells & \shortstack[r]{largest difference,\\agreeing cells} & \shortstack[r]{attributed\\to the table} \\
\midrule
\citet{tietjen1976variances} & normal & covariances, means & 5 to 49 & 239 & $9.88 \times 10^{-8}$ & 33 \\
\midrule
\citet{harter1960tables} & normal & mean of the range, variance of the range & 5 to 97 & 48 & $4.86 \times 10^{-11}$ & 0 \\
\midrule
\citet{teichroew1956tables}, Table I & normal & means & 2 to 20 & 100 & $4.91 \times 10^{-11}$ & 1 \\
\midrule
\citet{teichroew1956tables}, Table II & normal & products & 10, 20 & 43 & $4.96 \times 10^{-11}$ & 0 \\
\midrule
\citet{sarhan1956estimation} & normal & covariances & 5, 10, 20 & 80 & $8.70 \times 10^{-11}$ & 1 \\
\midrule
\citet{lieblein1953exact} & Gumbel & means, covariances & 3 & 9 & $4.38 \times 10^{-6}$ & 0 \\
\midrule
\citet{gupta1960order} & gamma & means, variances & 5, 10 & 30 & $8.56 \times 10^{-6}$ & 14 \\
\midrule
\citet{gupta1965exact} & logistic & covariances, means, second moments & 2, 5 & 7 & $8.88 \times 10^{-16}$ & 0 \\
\bottomrule
\end{tabular}
 \caption{\label{tab:tables} The printed tables compared with the package. A cell agrees
when the package's value lies within the accuracy its source states, $10^{-7}$ for Tietjen,
Kahaner, and Beckman, and one unit of the last printed digit for the others. A disagreeing
cell is attributed to the table when the conditional-beta check agrees with the package
within $10^{-10}$. The transcriptions, with their pages and conventions, are supplied with
the replication materials.}
\end{table}

Two tables carry errors far beyond the accuracy they state. In the table of
\citet{tietjen1976variances}, \cpTietjenGross{} cells at the quartile ranks of
$n \ge \cpTietjenFailNmin$ are off by up to \cpTietjenWorstFail, which is
\cpTietjenUnitsMax{} units of the $10^{-7}$ that the table certifies. In the table of
\citet{gupta1960order} for the gamma distribution with shape two, \cpGuptaTenFail{} of the
\cpGuptaTenCells{} cells at $n = 10$ disagree with the package, by \cpGuptaTenUnitsMin{} to
\cpExactTenUnitsMax{} units of the sixth significant figure. For this family, every moment
is a finite sum of rational numbers, so we also summed it in exact rational arithmetic,
which needs no floating point at all. The rational values agree with the package within
\cpExactPkgMax{} on all \cpExactCells{} cells, and they confirm the printed errors. The
other disagreements are small: they lie within \cpFineUnitsMax{} units of the criterion of
their source. Appendix~\ref{app:tables} discusses the disagreements of each table and their
possible causes.

\subsection{The other built-in families} \label{sec:newfamilies}

Beyond the five distributions mentioned above, the other eight built-in families are
Student's $t$, Cauchy, gamma, chi-square, beta,
lognormal, Weibull, and $F$. Except for the
gamma distribution with shape 2, which Section~\ref{sec:tables} compares with a printed
table, we found no printed table for them, and only special cases of them have closed
forms. We therefore check them against a \emph{reference}: the mean and the variance of
each order statistic computed by an integral in the probability variable $u = F(x)$ against
the beta density of $U_{(r)}$, as in the conditional-beta check of
Section~\ref{sec:tables}. This integral uses none of the package's integrands. At
\fmCheckCases{} parameter settings and the five-number ranks of $n = 5$, 25, 101, 501, and
1001, the package agrees with the reference within $\fmCheckPkgRefWorst$ on all
\fmCheckValues{} means and variances. Here a variance is compared in relative terms, and a
mean on the scale of its own order statistic (Appendix~\ref{app:newfamilies}). The
permutation identities hold within $\fmIdentWorst$ on \fmIdentMatrices{} full matrices, and
the covariances agree with the conditional-beta check on the \fmCovPairs{} pairs of ranks
for which that check converges (Table~\ref{tab:newfamilies}).

Two parameter regions are harder for the quadrature. The first region is a density that is
infinite at an endpoint of the support, which the change of variable of
Section~\ref{sec:devices} addresses. The second region is a variance that barely exists,
which means that the tail index times the rank is only slightly above the two that the
moment condition of Section~\ref{sec:closed} requires, as for the minimum of a sample from
Student's $t$ distribution with 2.001 degrees of freedom. In the two regions we
ran \fmRegionCells{} order statistics of \fmRegionCases{} parameter settings, at sample
sizes up to $n = 10^6 + 1$. The package returns the mean and the variance of
\fmRegionReturned{} of them, each within $\fmRegionPkgWorst$ of the reference in the same
terms, and it stops with an error for the remaining \fmRegionErrors. The covariances are
harder in these two regions. On full matrices at $n = \fmMatN$, \fmMatComputed{} of
\fmMatCases{} parameter settings are computed, and \fmMatErrors{} stop with an error rather
than return a number (Appendix~\ref{app:newfamilies}).

\section{Comparison with other software and computing time} \label{sec:comparison}

This section compares the package with the other tools that compute or approximate the
moments of order statistics, and then reports the computing time of the package.
Appendix~\ref{app:comparison} gives the details of both.

\subsection{Comparison with other software} \label{sec:tools}

We compare the package with every tool we found and could run, first by what each tool
computes, and then by how accurately it computes the quantities that it shares with the
package. No tool compared here returns a covariance between order statistics by exact
integration. Also note that \pkg{SciPy} can return a number for a moment that does not
exist and an inaccurate variance in a heavy tail, as the two examples of
Section~\ref{sec:intro} show. A third example is a density that is infinite at both
endpoints. For the maximum of 1001 observations from the $\mathrm{Beta}(1/2, 1/2)$
distribution, the variance that \pkg{SciPy} returns is \fmScipyArcsinePct\% away from the
reference of Section~\ref{sec:newfamilies}, and the package agrees with the reference
within $\fmPkgArcsineDiff$ in relative terms (Appendix~\ref{app:newfamilies}).

For the shared quantities, we compare the package, quantity by quantity, with each tool. Each
tool is run as its documentation describes and within its documented range, at sample
sizes up to $n = 1001$, as shown in Table~\ref{tab:tools}. In short, the tools
that integrate or use a closed form agree with the package until their own computation
breaks down at large $n$, and the approximate tools carry errors of up to
\cpMetaBlueCovRel\%.

\begin{table}[t!]
\centering
\footnotesize
\setlength{\tabcolsep}{3.5pt}
\begin{tabular}{llllrrrr}
\toprule
tool & family & quantities & method & \shortstack[r]{sound\\to $n$} & \shortstack[r]{first\\failing $n$} & \shortstack[r]{largest\\difference} & \shortstack[r]{relative\\(\%)} \\
\midrule
\pkg{EnvStats} & normal & means & exact & 1001 & -- & $1.13 \times 10^{-13}$ & -- \\
\pkg{SuppDists} & normal & means & approximate & -- & -- & $7.92 \times 10^{-5}$ & 0.0 \\
\pkg{lmomco} & normal & means & exact & 101 & 501 & $9.77 \times 10^{-7}$ & -- \\
\pkg{lmomco} & exponential & means & exact & 101 & 501 & $4.89 \times 10^{-7}$ & -- \\
\pkg{lmomco} & Gumbel & means & exact & 101 & 501 & $5.32 \times 10^{-7}$ & -- \\
\pkg{mos} & uniform & means & exact & 1001 & -- & $1.02 \times 10^{-13}$ & -- \\
\pkg{mos} & uniform & variances & exact & 1001 & -- & $1.97 \times 10^{-13}$ & -- \\
\pkg{mos} & exponential & means & exact & 25 & 101 & $6.68 \times 10^{-8}$ & -- \\
\pkg{mos} & exponential & variances & exact & 25 & 101 & $1.79 \times 10^{-7}$ & -- \\
\pkg{mos} & normal & means & approximate & -- & -- & $2.17 \times 10^{-3}$ & 0.2 \\
\pkg{mos} & normal & variances & approximate & -- & -- & $3.16 \times 10^{-3}$ & 1.0 \\
\pkg{metaBLUE} & normal & means & approximate & -- & -- & $3.66 \times 10^{-3}$ & 0.3 \\
\pkg{metaBLUE} & normal & covariance matrix $\bB$ & approximate & -- & -- & $2.67 \times 10^{-2}$ & 6.0 \\
\pkg{SciPy} & normal & means, variances & exact & 1001 & -- & $5.38 \times 10^{-13}$ & -- \\
\pkg{SciPy} & uniform & means, variances & exact & 1001 & -- & $6.02 \times 10^{-13}$ & -- \\
\pkg{SciPy} & exponential & means, variances & exact & 1001 & -- & $1.08 \times 10^{-12}$ & -- \\
\pkg{SciPy} & logistic & means, variances & exact & 1001 & -- & $9.06 \times 10^{-13}$ & -- \\
\pkg{SciPy} & Gumbel & means, variances & exact & 1001 & -- & $6.41 \times 10^{-10}$ & -- \\
\bottomrule
\end{tabular}
 \caption{\label{tab:tools} The other tools against the package, by tool, family, and
quantity. For a tool that integrates or uses a closed form, the largest $n$ up to which
every sample size run agrees with the package within $10^{-6}$ (its sound range), the first
$n$ at which it fails or returns a non-finite value, and the largest absolute difference
within the sound range. For an approximate tool, the largest absolute and relative
differences over every sample size run, with the relative difference defined in
Appendix~\ref{app:tools}. \pkg{metaBLUE} is compared on the mean vector and the covariance
matrix of the five-number summary.}
\end{table}

Among the tools that integrate or use a closed form, \pkg{EnvStats} and \pkg{SciPy} agree
with the package at every sample size on the families of Table~\ref{tab:tools}. \pkg{lmomco} agrees up to $n = \cpLmomcoSoundTo$ and
loses its accuracy from $n = \cpLmomcoFirstFail$ on. The closed forms of \pkg{mos} for the
exponential distribution agree up to $n = \cpMosExpSoundTo$, and at $n = \cpMosExpFirstFail$
they return values with absolute errors as large as \cpMosExactWorst. Among the approximate
tools, the series of \pkg{metaBLUE} misses the entries of $\bB$ by up to
\cpMetaBlueCovRel\%, and its variances of the minimum and the maximum are off by
\cpMetaBlueExtVarRelAtFive\% at $n = 5$, \cpMetaBlueExtVarRelAtHundredOne\% at $n = 101$,
and \cpMetaBlueExtVarRelAtThousand\% at $n = 1001$. Appendix~\ref{app:tools} discusses each
tool.

\subsection{Computing time} \label{sec:timing} 

The cost of one integral hardly depends on $n$. For the standard normal distribution at the
five-number ranks, a single integral costs \atTimePerSingleAtMin{} ms at $n = \atTimeNmin$
and \atTimePerSingleAtMax{} ms at $n = \atTimeNmax$. A double integral costs
\atTimePerDoubleAtMin{} and \atTimePerDoubleAtMax{} ms at the same two sample sizes, a
growth by a factor of \atTimeDoubleGrowth{} (Appendix~\ref{app:timing},
Table~\ref{tab:timing}, and Figure~\ref{fig:timing}). In a call with
covariances at these ranks, the double integrals take
\ifnum\atTimeShareDoubleMin=\atTimeShareDoubleMax about \atTimeShareDoubleMax\%\else
\atTimeShareDoubleMin\% to \atTimeShareDoubleMax\%\fi{} of the elapsed time, because every
evaluation point of an outer integral runs an inner adaptive integration. The cost of a call
is therefore the number of pairs times a few tens of milliseconds. A full matrix needs one
double integral per pair, so its cost grows with the square of $n$. A full normal matrix
took \atIdentTimeHundredMin{} minutes at $n = 100$ on one core, and \atLargeWallMin{}
minutes at $n = \atLargeN$ on \atLargeCores{} cores as a parallel job over pairs. The package versions and
the computer used for these times are given in the Computational details section below.

\Needspace*{4\baselineskip}
\section{Summary and discussion} \label{sec:summary}

Our proposed package \pkg{orderStatMoments} computes the exact means, variances, and
covariances of order statistics for any ranks and any sample size, for thirteen built-in
families, specified as in \proglang{R}, and for a continuous distribution that the user
supplies through its distribution, density, and quantile functions. It uses closed forms where they exist. On top of the
moments, it computes the moments of any linear
combination of order statistics, such as sample quantiles and Lloyd's BLUEs. Its two contributions are the exact covariances, which no tool we found
computes, and the design of the computing methods in Section~\ref{sec:devices}, which we built on
the statistical properties of order statistics.

As described in Section~\ref{sec:accuracy}, the evidence of accuracy covers all thirteen
built-in families: for every family, the package agrees with at least one independent
source of values. The first source is the closed forms of four families, which the
quadrature reproduces within \atClosedWorst{} at sample sizes up to
$n = \atClosedSizeMax$. The second source is the printed tables of
\countword{\cpPublicationsTotal} publications, the normal tables among them. The package
agrees with \cpCellsAgree{} of their \cpCellsTotal{} cells, and on each of the other
\cpCellsTable{} cells the conditional-beta check of Section~\ref{sec:tables} sides with the
package. The third source, for the other eight families, is an integral in the
probability variable that uses none of the package's integrands, with which the means and
the variances agree within $\fmCheckPkgRefWorst$ in relative terms. In addition, the
permutation identities hold within $10^{-9}$ on every full covariance matrix that the
package returned, for the families that have a variance.

The package has its own limitations.
A user-supplied distribution must honor \code{log.p}
and \code{lower.tail}. Its tail indices, its symmetry, and the points where its density is
not smooth (Appendix~\ref{app:laplace}) should be declared by the user. Two
numerical limits remain, and each ends in an error or shows in the run-time
checks. First, a user-supplied distribution whose density is infinite at an upper endpoint
has no change of variable, so the order statistics next to that endpoint are computed less
accurately or refused. Supplying the reflected variable, whose density is infinite at the
lower endpoint, removes this limit. Second, a moment can exist and
still be out of reach of the quadrature. An example is the covariance between the first and
the fourth order statistics of $n = 20$ observations from Student's $t$ distribution with
2.1 degrees of freedom, for which the package stops with an error. In this example the
variance of the first order statistic barely exists, in the sense of
Section~\ref{sec:newfamilies}, and the quadrature does not resolve the heavy lower tail of
the product moment. The mean and the variance of a single order statistic of the same
distribution are within reach, as the second example of Section~\ref{sec:intro} shows. Also, our accuracy
checks are broad, but they cannot cover every case. The help page of \code{os_moments()}
lists the cases that we tested. The most difficult extreme cases, as well as more built-in
families, will be addressed in future development.

\section*{Computational details}

The results in this paper were obtained with \rpRVersion{} on an \atTimeMachine. The code
chunks and every study ran \pkg{orderStatMoments} \rpPkgVersion. The tools of
Section~\ref{sec:comparison} are \cpToolVersions{}, and \pkg{SciPy} \cpScipyVersion. Every
number quoted from the accuracy and comparison studies, every table and figure, and every
code chunk is read from the outputs of the replication scripts, which run the package on the
sample sizes and ranks stated. The tolerances named in Section~\ref{sec:computation} are the
package's constants. \proglang{R} and the \proglang{R} packages used are available from the
Comprehensive \proglang{R} Archive Network (CRAN) at \url{https://CRAN.R-project.org/},
except \pkg{orderStatMoments}. It is available from
\url{https://github.com/ZWuLab/orderStatMoments}, with its documentation at
\url{https://zwulab.github.io/orderStatMoments/}. The \proglang{Python} side ran
\proglang{Python} \cpPythonVersion{} with \pkg{SciPy} \cpScipyVersion{} and \pkg{NumPy}
\cpNumpyVersion. 

\section*{Acknowledgments}

Zheyang Wu is supported in part by NSF grant MMS-2515791.

\Needspace*{6\baselineskip}   %

\ifseparateappendix\else
\newpage

\begin{appendix}

\setcounter{table}{0}
\renewcommand{\thetable}{A\arabic{table}}
\renewcommand{\theHtable}{A\arabic{table}}
\setcounter{figure}{0}
\renewcommand{\thefigure}{A\arabic{figure}}
\renewcommand{\theHfigure}{A\arabic{figure}}

\section{Background on the integrands and the integrator} \label{app:background}

This appendix gives the background behind Section~\ref{sec:background} in more detail: how
the width of an order statistic behaves as $n$ grows, how \code{integrate()} proceeds, and
which limits a double-precision number has.

\subsection{The width of an order statistic at central and extreme ranks} \label{app:width}

How the width $s_r/f(c_r)$ in \eqref{eq:width} behaves as $n$ grows depends on the rank,
in two cases \citep[Section~10.1]{david2003order}. For a \emph{central rank}, where
$r/n \to p$ with $0 < p < 1$, $s_r \approx \sqrt{p(1-p)/n}$, so the width shrinks like
$n^{-1/2}$. In this case $\sqrt{n}\{X_{(r)} - Q(p)\}$ is asymptotically normal with
variance $p(1-p)/f(Q(p))^2$, provided $r/n - p = o(n^{-1/2})$ and $0 < f(Q(p)) < \infty$
\citep[Theorem~10.3]{david2003order}. This asymptotic variance is $n$ times the square of
the width in \eqref{eq:width}.

For an \emph{extreme rank}, where $r$ or $n - r + 1$ stays fixed,
$s_r \approx \sqrt{r}/n$ for a fixed $r$ and $s_r \approx \sqrt{n - r + 1}/n$ for a fixed
$n - r + 1$. The center $c_r$ moves into the lower tail or the upper tail, so the width is
governed by $f(c_r)$. The width of the minimum is about $1/n$ for the uniform distribution
and of order $1/\sqrt{\log n}$ for the normal distribution. The width of the maximum stays
of order one for the exponential, logistic, and Gumbel distributions, where the variance of
the maximum tends to $\pi^2/6$ (Section~\ref{sec:closed}). The width grows with $n$ for a
heavy-tailed parent distribution.

\subsection[How integrate() works]{How \code{integrate()} works} \label{app:integrate}

\proglang{R}'s \code{integrate()} function applies the QUADPACK routine \code{dqags} on a
finite range and \code{dqagi} on an infinite range
\citep{piessens1983quadpack, rcoreteam2023r}. Its basic step is a Gauss--Kronrod rule,
applied to a given interval. The rule evaluates the integrand at 21 fixed points of the
interval, called the nodes, and approximates the integral over the interval by a weighted
sum of the 21 values. To estimate the error, ten of the nodes form a Gauss rule on their
own, and the difference between the two weighted sums is the basis of the error estimate
of the step.

The calculation procedure is adaptive. The routine first applies the rule to the whole
range, which gives one value and one error estimate. It then bisects the range, applies
the rule to each half, and keeps the two halves, with their values and error estimates, in
place of the whole. From then on, each step takes the subinterval with the largest error
estimate among those kept, bisects it, applies the rule to the two halves, and replaces
the subinterval by them. At any step the current value $I$ of the integral is the sum of
the values over the subintervals kept, and the error estimate $E$ is the sum of theirs
(the routine also extrapolates from the sequence of values, a refinement we leave aside).
It stops when the number of subintervals reaches its limit, or when $E$ satisfies the
stopping rule \eqref{eq:stop}. An infinite range $(a, \infty)$ is mapped onto $(0, 1]$ by
$x = a + (1 - t)/t$, and a 15-point rule of the same kind is applied in $t$. The doubly
infinite range $(-\infty, \infty)$ is split at zero.

A double integral is computed as a nested quadrature: an outer integral over $y$ whose
integrand is itself an inner integral over $z$, which is computed by one call of
\code{integrate()} at each outer node. \code{integrate()} returns the value, the error
estimate, and a message. The message is \code{"OK"}, or it names a problem: the limit on the
subintervals was reached, roundoff error was detected, the integrand behaves badly, or the
integral is probably divergent.

The help page of \code{integrate()} illustrates the blind spot of
Section~\ref{sec:background} in its Examples: \code{integrate(dnorm, 0, 20000)} wrongly
returns 0, because the integrand is close to zero over nearly all its range.

\subsection{Double-precision numbers} \label{app:double}

\proglang{R} stores a number as a double-precision floating-point number, called a double. A
double carries about sixteen significant decimal digits: the relative spacing of the
representable numbers is $\epsilon = 2^{-52} \approx 2.2 \times 10^{-16}$. The largest
double is about $1.8 \times 10^{308}$, and a result beyond it overflows to infinity. The
smallest positive double of full precision is about $2.2 \times 10^{-308}$, and a positive
result below it is stored with reduced precision and then underflows to zero.

The three challenges of Section~\ref{sec:background} follow from these limits. For the
first, the overflow or underflow of a product of many factors, the solution is to work with
the logarithm of the product (Appendix~\ref{app:log}). For the second, the cancellation,
if $a$ and $b$ agree in their first $k$ significant digits, $a - b$ keeps accuracy in about
$16 - k$ digits. For the third, a probability very close to one is stored with an absolute
error of about $\epsilon$, so $1 - F(x)$ formed from a stored $F(x)$ has a relative error
of about $\epsilon / \{1 - F(x)\}$. Taking the upper tail probability directly, and its
logarithm, also keeps a tiny tail probability from underflowing. The solutions to the second
and the third are the moments taken about the center of each order statistic
(Appendix~\ref{app:cancel}) and the difference formed on the side where the probabilities
are small (Appendix~\ref{app:diff}).

\section{Details of the numerical design} \label{app:design}

This appendix gives the details of the solutions in Section~\ref{sec:devices}, under the
same headings. Table~\ref{tab:challenges} lists every computing challenge that we met,
the cases in which it arises, and the solution that resolves it. The subsections then
treat the solutions in the order of the table, from its second row on, and the last
subsection gives the details
of Section~\ref{sec:closed}.

\begin{table}[t!]
\centering
\small
\begin{tabular}{>{\raggedright\arraybackslash}p{4.4cm} >{\raggedright\arraybackslash}p{3.7cm} >{\raggedright\arraybackslash}p{5.5cm}}
\toprule
Challenge & Where it bites & Solution \\
\midrule
The fixed interval lengths of the integration and the absolute tolerances & a parent distribution for which these lengths are too wide or too tight & the standardized units \eqref{eq:units} \\
\midrule
Cancellation in $\Ex[X_{(r)}^2] - \mu_{r:n}^2$ & large $n$ & each moment about its own center $\tilde c_r$, \eqref{eq:centered} and \eqref{eq:assemble} \\
\midrule
The nodes of the rule miss the peak of the integrand & large $n$, at every rank & the range cut at $\tilde c_r \pm h_r$, \eqref{eq:split} \\
\midrule
The integrand of a first moment changes sign at $\tilde c_r$ & the first moments & the range cut at $\tilde c_r$ as well \\
\midrule
The noise of the inner quadrature seen by the outer rule & every product moment & the outer relative tolerance set to a hundred times the inner one \\
\midrule
A moment smaller than a fixed absolute tolerance & sharp order statistics, such as the minimum or the maximum of a bounded parent distribution at large $n$ & the absolute tolerances scaled by the width of the order statistic \\
\midrule
The outer factor magnifies the inner error & the far tail of a heavy-tailed parent distribution & the inner tolerance divided by $\max(1, |y - \tilde c_r|)$ \\
\midrule
A wrong value but with a small error estimate and no error message & a missed mode, a moment that does not exist, a variance returned as zero or negative & the refusal rules: the message rule, the divergence rule, the mass check, and the variance guard \\
\midrule
A density that is infinite at an endpoint of the support & a gamma, a Weibull, or a beta distribution with a shape parameter below one, a chi-square distribution or the numerator of an $F$ distribution with fewer than two degrees of freedom & the pieces next to the endpoint integrated in a power of the distance from it (Appendix~\ref{app:endpoint}) \\
\midrule
Overflow of the coefficient, underflow of the powers & large $n$ & the integrand evaluated on the logarithmic scale, \eqref{eq:logint} \\
\midrule
Cancellation in $F(z) - F(y)$ & both points in one tail & the difference formed on the side where the probabilities are small \\
\bottomrule
\end{tabular}
\caption{\label{tab:challenges} The computing challenges, where each one bites, and the
design of Section~\ref{sec:devices} that resolves it, in the order of the text.}
\end{table}

\subsection{Formulating moments: the size of the cancellation} \label{app:cancel}

By \eqref{eq:cancel}, a variance that is computed as $\Ex[T_{(r)}^2] - (\Ex[T_{(r)}])^2$
has a relative error of about $\epsilon_q\,\tilde c_r^2/\Var(T_{(r)})$, where $\epsilon_q$
is the relative tolerance of each integral. Here we give the calculation of this ratio that
Section~\ref{sec:devices} quotes.

This ratio grows with $n$. For example, at a central rank, $\tilde c_r$ is fixed while
$\Var(T_{(r)})$ shrinks at the rate $1/n$ by \eqref{eq:width}, so the ratio grows at the
rate $n$. In the more extreme case of the minimum or the maximum of a parent distribution
whose density is positive and finite at a finite endpoint of its support, the ratio can grow
like $n^2$. For example, for the uniform parent distribution, whose median is $1/2$, the
minimum has center $1/(n+1)$ and variance $n/\{(n+1)^2(n+2)\}$ in the original units, so
\begin{equation*}
\frac{\tilde c_1^2}{\Var(T_{(1)})}
= \frac{\{1/(n+1) - 1/2\}^2/\tilde\sigma^2}{n/\{(n+1)^2 (n+2)\,\tilde\sigma^2\}}
= \frac{\{1/2 - 1/(n+1)\}^2}{n/\{(n+1)^2 (n+2)\}} \approx \frac{n^2}{4},
\end{equation*}
which is $2.5 \times 10^5$ at $n = 10^3$ and $2.5 \times 10^9$ at $n = 10^5$. With
$\epsilon_q = 10^{-11}$, the relative error of the variance is about $2.5 \times 10^{-6}$
and $2.5 \times 10^{-2}$, respectively. By this estimate, the variance keeps only about five
accurate digits at the first sample size and fewer than two at the second. Once the ratio
reaches $1/\epsilon_q$, which the uniform minimum does at about $n = 6 \times 10^5$, the
variance is \emph{lost} in the sense that the returned value can be zero or even negative. A
covariance can be lost in the same way, with the ratio $\tilde c_r \tilde c_s/\Cov(T_{(r)},
T_{(s)})$ in place of the ratio in \eqref{eq:cancel}. The risk is the largest for the
covariance of two nearly independent order statistics, which is small.

\subsection{Creating integration splits around the peaks} \label{app:split}

This subsection gives three details of the cuts in Section~\ref{sec:devices}: why the outer
pieces of a single integral are safe, why the outer integral of a product moment is
cut at the points of both ranks, and how a density that is not smooth is handled.

For the second moment $v_r$ in \eqref{eq:centered}, the peak of $f^T_{r:n}$ sits inside the
finite middle piece, and the two outer pieces carry the tails. For a parent distribution
with a log-concave density, which includes the normal, uniform, exponential, logistic, and
Gumbel families, $f^T_{r:n}$ is
log-concave as well. It therefore decreases away from its peak, and the integrand $(t -
\tilde c_r)^2 f^T_{r:n}(t)$ of $v_r$ has at most one local maximum on each side of $\tilde
c_r$, so neither outer piece hides a peak between the nodes. For other parent distributions,
for example one with a second mode, an outer piece can hide a peak, and the mass check of
Appendix~\ref{app:rules} is the safeguard.

The product moment $v_{rs}$ in \eqref{eq:centered} is computed as the nested quadrature
\begin{equation*}
v_{rs} = \int_{t_L}^{t_U} (y - \tilde c_r)\, g(y)\, dy, \qquad
g(y) = \int_y^{t_U} (z - \tilde c_s)\, f^T_{r,s:n}(y, z)\, dz,
\end{equation*}
where $(t_L, t_U)$ denotes the support of $T$, and each outer node $y$ calls
\code{integrate()} once for computing $g(y)$. We cut the range $(y, t_U)$ of the inner
integral at those of the points $\tilde c_s - h_s$, $\tilde c_s$, and $\tilde c_s + h_s$
that lie inside it, and we cut the range of the outer integral at the same three points to
match.
This matching is needed because as $y$ crosses $\tilde c_s - h_s$, $\tilde c_s$, or
$\tilde c_s + h_s$, the inner integral loses the corresponding piece, and its computed value
$g(y)$ can change by a small amount, about the size of the inner error. The outer integrand
is therefore non-smooth at these points, a feature that the adaptive rule of
\code{integrate()} does not accommodate well. Section~\ref{sec:identities} reports an
example where one product moment went wrong when the outer integral was cut at the points of
rank $r$ alone.

The cuts above assume a density that is smooth inside the support, as the densities of the
built-in families are. For a user-supplied distribution whose density has a kink or a jump,
the user declares these points in the argument \code{breaks} of \code{os_dist()}. We add
them to the cut points of every single integral, and of the inner and the outer integral of
every product moment, so that the integrand is smooth inside each piece
(Appendix~\ref{app:laplace}).

\subsection{Setting proper tolerances} \label{app:tol}

We set the two tolerances in \eqref{eq:stop}
for each integral, in the standardized units. For the relative tolerance, we set it to
$10^{-11}$ for the single integrals and the inner integral of a product moment, and to
$10^{-9}$ for the outer integral. The outer integrand is itself a quadrature value, so it
carries the much smaller inner error as noise, which the outer rule need not chase.

By \eqref{eq:stop}, an absolute tolerance matters only when it exceeds
$\mathrm{rel.tol} \times |I|$. We set the absolute tolerances to $10^{-13}$ for a single
integral, $10^{-15}$ for an inner integral, and $10^{-12}$ for an outer integral. The inner
one is the tightest because the inner value is multiplied by the outer factor and integrated
again. The outer one is the loosest because its integrand carries the inner noise.

A fixed absolute tolerance can exceed the moment it is meant to compute. In particular, the
order statistics of a bounded parent distribution at large $n$ are sharp. For example, the
variance of the minimum of $10^6$ uniform observations is $7 \times 10^{-12}$ in
standardized units. A quadrature that stops at an absolute error of $10^{-13}$ may then be
off by $1.4\%$ of the value, which leaves only its first two significant digits reliable.
Our solution is to multiply each absolute tolerance by $\min(1, d_r)$ if $d_r$ in
\eqref{eq:split} is nonzero. This is done once for each factor $(t - \tilde c_r)$ or $(z -
\tilde c_s)$ in the integrand, because a moment with $k$ such factors is of the order of the
$k$th power of the width, where $k = 1$ or $2$ is the order of the moment. The factor is
$\min(1, d_r)$ for $m_r$, $\min(1, d_r)^2$ for $v_r$, $\min(1, d_s)$ for the inner integral
of $v_{rs}$, and $\min(1, d_r)\times \min(1, d_s)$ for its outer integral. The tolerances
thus tighten with a sharp order statistic and never loosen beyond their constants.

The inner integral of $v_{rs}$ is multiplied by the outer factor $y - \tilde c_r$ before it
is integrated over $y$. In the far tail of a heavy-tailed parent distribution, $|y - \tilde
c_r|$ is large, because the adaptive rule places nodes far out where the tail decays slowly.
For example, for a lognormal parent distribution with a log-scale standard deviation of 2,
at $n = 2$ and the ranks 1 and 2, the outer nodes reach $|y - \tilde c_r|$ of
$\rpLognormOuterReach$ in standardized units. An inner error of $10^{-15}$ is multiplied by
$|y - \tilde c_r|$ in the product. We therefore divide the inner absolute tolerance by
$\max(1, |y - \tilde c_r|)$ as well.

\subsection{Setting proper acceptance and refusing rules} \label{app:rules}

As discussed in Section~\ref{sec:background}, \code{integrate()} may return a wrong value
with a small error estimate and no warning message when it computes \eqref{eq:moments}, and
a warning message may not indicate a wrong value. Therefore, we set the following rules for
accepting or refusing a result based on solid evidence and proper checks, not just on the
messages.
\begin{enumerate}
\item \emph{The tail probe.} Before any integration, the log-tails and the log-density of
  the parent distribution are evaluated at the cut points \eqref{eq:split} of the smallest
  and the largest requested rank and at the median. A value that is not finite will stop the
  computation, for example when the support of a bounded parent distribution was not
  declared. A user-defined distribution function that ignores \code{log.p} or
  \code{lower.tail} is caught earlier, when \code{os_dist()} builds the distribution
  (Section~\ref{sec:userdist}).
\item \emph{The message rule.} A piece whose message is not \code{"OK"} may still be
  accepted when its error estimate is below a threshold. The threshold is $10^{-10}$
  multiplied by the same factors $\min(1, d_r)$ and $\min(1, d_s)$ that multiply the
  absolute tolerance of the integral (Appendix~\ref{app:tol}), so that the threshold also
  tightens for a sharp order statistic. Otherwise the computation stops with an error
  message. A usual harmless case is roundoff error reported on a tail piece whose value is
  near zero, where the relative tolerance cannot be met.
\item \emph{The divergence rule.} A piece on an infinite range that \code{integrate()}
  reports as probably divergent may still be accepted when it passes the message rule and,
  in addition, the absolute value of the piece is below the threshold of the message rule,
  that is, $10^{-10}$ multiplied by the same factors. The reason is that a second moment
  that does not exist can come back as a finite value of ordinary size with a small error
  estimate, while a negligible tail piece of a convergent integral can carry the same
  message. For example, the minimum of five observations from a $t$ distribution with 1.9
  degrees of freedom has no second moment. With the tail indices left undeclared,
  \code{integrate()} reports the lower tail piece of this second moment as probably
  divergent, and this rule stops the computation. On a finite piece the integral cannot
  diverge, because the integrand is a power of $t$ times a density, which is integrable on a
  bounded interval. We therefore treat the message as spurious there and judge the piece by
  its error estimate like any other.
\item \emph{The mass check.} The density $f^T_{r:n}$ of every order statistic whose mean is
  computed by quadrature is integrated with the same cuts, at the cost of one more single
  integral per rank, and the computation stops when the mass differs from one by more than
  $10^{-7}$. This check is needed because a second mode of the parent distribution can fall
  between the nodes of a piece. For the mixture $0.9\,N(0, 0.01^2) + 0.1\,N(10, 0.01^2)$,
  the density of the median of five observations integrates to \rpMixMass{}, because the
  mode at 10 fell between the nodes. Without this check, the moments returned would be those
  of the mode at 0 alone, and the identities of Section~\ref{sec:closed} do not catch this
  error, because the moments of the parent distribution that they compare against are
  computed by the same quadrature and miss the same mode. The threshold $10^{-7}$ is loose
  against the accuracy of the quadrature, so that a small inaccuracy of a user's tail
  functions does not stop the computation, while a missed mode, which loses a visible
  fraction of the mass, does.
\item \emph{The variance guard.} The variance of a continuous order statistic is positive,
  so a variance that comes out zero or negative signals that the quadrature lost it
  (Section~\ref{sec:devices}). Such a variance stops the computation, with an error message
  that names the rank.
\end{enumerate}

\subsection{Integrating next to an endpoint where the density is infinite} \label{app:endpoint}

This appendix gives the details of the change of variable of Section~\ref{sec:devices}: why
the standardized units fail next to an endpoint where the density of the parent distribution
is infinite, the change of variable, how the range is cut, and how the result is assembled.

Suppose that the support of $X$ has a finite lower endpoint $L$, so that the support of $T$
has the lower endpoint $\ell = \{L - Q(1/2)\}/\tilde\sigma$, and write $\delta = t - \ell$
for the distance from it. Suppose also that $f_T(\ell + \delta)\,\delta^{1 - \theta}$ tends to
$\theta C$ as $\delta \to 0$, with a constant $C > 0$ and an index $0 < \theta < 1$. The density
is then infinite at $\ell$, the ratio of $F_T(\ell + \delta)$ to $C \delta^{\theta}$ tends to
one, and by \eqref{eq:single} the density of $T_{(r)}$ is of the order of
$\delta^{\theta r - 1}$. The index $\theta$ is the shape parameter of the gamma and
of the Weibull distribution, half the degrees of freedom of the chi-square distribution and
of the numerator of the $F$ distribution, and the first shape parameter of the beta
distribution, whose second shape parameter plays the same role at the upper endpoint.

Two things can fail in the standardized units. First, \code{integrate()} converges slowly at
such a singularity, and at our tolerances it stops with a warning message, which the message
rule of Appendix~\ref{app:rules} turns into an error. Second, an order statistic at an
extreme rank can lie too close to the endpoint for a double-precision number in $t$ to tell
the two apart. For the
gamma distribution with shape 0.1, the center of the minimum of $n = 1001$ observations lies
$2 \times 10^{-29}$ from $\ell$ in units of the spread, while $\ell = -0.023$ itself is
stored to about $3 \times 10^{-18}$. In $t$, this center is the endpoint.

\paragraph{The change of variable.} On a piece $(\ell, \ell + \Delta)$ next to the endpoint,
we integrate in $\eta = \delta^{\theta}$:
\begin{equation*}
\int_{\ell}^{\ell + \Delta} g(t)\, dt = \int_0^{\Delta^{\theta}} g\bigl(\ell + \eta^{1/\theta}\bigr)\,
\frac{1}{\theta}\, \eta^{1/\theta - 1}\, d\eta .
\end{equation*}
In $\eta$, the distribution function is nearly linear, $F_T \approx C \eta$, and the density of
$T_{(r)}$ with the factor $\eta^{1/\theta - 1}/\theta$ is of the order of $\eta^{r - 1}$, which
is bounded at every rank. The functions of the parent distribution are evaluated at
$x = L + \tilde\sigma \delta$ with $\delta = \eta^{1/\theta}$, and not at
$Q(1/2) + \tilde\sigma t$. For
the families above, $L = 0$, so $x = \tilde\sigma \delta$ carries the full relative precision
of $\delta$. The deviation $t - \tilde c_r$ in \eqref{eq:centered} is computed as
$\delta - \delta_r$, where
$\delta_r = \{Q(p_r) - L\}/\tilde\sigma$ is the distance of the center from the endpoint, taken
from the quantile function directly. The quantity $d_r$ in \eqref{eq:split}, which scales the
tolerances (Appendix~\ref{app:tol}), takes the density at the center from the same distance.

\paragraph{The cuts.} A finite piece is integrated in $\eta$ in two cases. The first case
is that the distance from $\ell$ to
the nearer end of the piece is at most the length of the piece, so that the singularity is
felt inside it. The second case is that double precision does not resolve the piece in $t$.
We count two roundings, that of a node $t$ and that of the argument
$Q(1/2) + \tilde\sigma t$ at which the functions of the distribution are evaluated, and the
piece is resolved when both are below $10^{-12}$ of its length. Such a piece can lie farther
from $\ell$ than it is long and still be a few spacings of the doubles long in $t$. An example
is one of the pieces around the order statistic of rank 39,200 of $n = 10^6$ observations
from a beta distribution with the shape parameters $1/14$ and 1. The rounding of the argument
matters when the median is large against the spread. An example is the beta distribution
with the shape parameters 1 and $1/40$, whose median is within $10^{-12}$ of one while
$\tilde\sigma$ is below $10^{-5}$. The other pieces are
integrated in $t$ as before. Each cut
point carries its distance from the endpoint, and the cut points are ordered by these
distances, because the centers of two extreme order statistics can both equal $\ell$ in $t$.
The mass check of Appendix~\ref{app:rules} integrates the density of the order statistic
over the same pieces, in the same variables, as its two moments, so that a piece that double
precision does not resolve affects the mass as well as the moments.
The cut \eqref{eq:split} is not enough next to such an endpoint, because its lower bound
$4/\sqrt{n+1}$ on $h_r$ is far wider than an extreme order statistic there. Suppose that
the rank is extreme enough for $F_T \approx C\eta$ to hold over the range of $T_{(r)}$. Then
$(T_{(r)} - \ell)^{\theta}$ is close to $U_{(r)}/C$, so in $\eta$ the order statistic has its
center at $\eta_r = \delta_r^{\theta}$, and the ratio of its standard deviation to its mean is
about $s_r/p_r$, by \eqref{eq:beta}. We therefore add cuts at
\begin{equation*}
\eta_r \Bigl(1 - 10\, \frac{s_r}{p_r}\Bigr), \qquad \eta_r \Bigl(1 - 40\, \frac{s_r}{p_r}\Bigr),
\qquad \eta_r \Bigl(1 + 10 \cdot 4^{j}\, \frac{s_r}{p_r}\Bigr), \quad j = 0, 1, 2, \dots,
\end{equation*}
the first two where they are positive, and the others up to the upper cut point in
\eqref{eq:split}. The spacing of
the upper cuts grows by the factor four because the integrand of a moment of order $k$ is the
density times $\delta^{k} = \eta^{k/\theta}$, whose peak lies far above the peak of the density
when $\theta$ is small. For the minimum at a large $n$, this integrand is of the order of
$\eta^{k/\theta} e^{-\eta/\eta_r}$, which peaks near $(k/\theta)\,\eta_r$, that is, near
$40\,\eta_r$ for $\theta = 0.05$ and $k = 2$. In a
product moment, an outer node $y$ in such a piece is passed to the inner integral as its
distance, so that $F_T(z) - F_T(y)$ and the two deviations keep their precision.

\paragraph{The result.} The mean of an order statistic computed in this way is returned as
$L + \tilde\sigma (\delta_r + m_r)$, and not through the median as in
Section~\ref{sec:devices}, so that a mean of $10^{-24}$ keeps its relative precision.

\paragraph{The upper endpoint and a user-supplied distribution.} At a finite upper
endpoint, the same construction applies with $\delta$ the distance below the endpoint,
$1 - F_T$ in place of $F_T$, $n - r + 1$ in place of $r$, and $1 - p_r$ in place of $p_r$.
However, a point at the distance $\tilde\sigma\delta$ below an endpoint that is not zero
loses the digits of $\delta$ below the rounding of the endpoint. For the beta
distribution, $1 - X$ has the beta distribution with the two shape parameters exchanged, so
the package evaluates the functions of $1 - X$ at $\tilde\sigma \delta$ and loses nothing. For a
built-in family, $\theta$ is the family's own parameter. For a user-supplied distribution,
the package finds such an endpoint by an infinite value of the density there, and it
estimates $\theta$ from the slope of $\log F_T$ against $\log \delta$. With an estimated
index the integrand in $\eta$ can keep a weak singularity, which \code{integrate()}
resolves. The package has no functions of the reflected variable for an upper
endpoint of a user-supplied distribution, which it therefore integrates in $t$. The user
can supply the reflected variable in place of $X$, that is, the variable whose value is the
upper endpoint minus $X$. Its support starts at zero, and its order statistic of rank
$n + 1 - r$ is the upper endpoint minus $X_{(r)}$. Appendix~\ref{app:newfamilies} gives the
evidence for the change of variable.

\subsection{Computing at the logarithmic scale} \label{app:log}

This subsection gives the details of the computation of the integrand on the logarithmic
scale in Section~\ref{sec:devices}: where the direct product fails, how each term in
\eqref{eq:logint} is obtained, and what the logarithmic scale costs.

The coefficient $n!/\{(r-1)!\,(n-r)!\}$ in \eqref{eq:single}, when it is computed from its
factorials, overflows from $n = 171$ on, because $171!$ exceeds the largest double. Even
when the coefficient is written as the binomial expression $n\binom{n-1}{r-1}$, it
overflows at a central rank from $n = 1021$ on.

The logarithm of the coefficient in \eqref{eq:logint} is obtained from \code{lbeta()}
rather than from a difference of \code{lgamma()} values, which can lose precision by
cancellation at large $n$, e.g., in the millions. For the joint density, the coefficient is
$1/\{\mathrm{B}(r, n-r+1)\,\mathrm{B}(s-r, n-s+1)\}$. The logarithms of the two tail
probabilities are obtained from the distribution function, by \code{p(x, log.p = TRUE)} and
\code{p(x, lower.tail = FALSE, log.p = TRUE)}, where \code{p} stands for \code{pnorm()}
for the normal distribution and its analogues for the other distributions. These two
logarithms keep their relative precision where the probabilities themselves would
underflow, as \proglang{R}'s distribution functions do and as we require of a user's
functions (Section~\ref{sec:userdist}). The log-density is obtained from
\code{d(x, log = TRUE)}.

The price of the logarithmic scale is that the exponential of a stored logarithm $\ell$
carries a relative error of about $\epsilon|\ell|$, which stays below $2 \times 10^{-13}$
for $|\ell| \le 700$, roughly the range in which $e^{\ell}$ can be stored as a double. In
addition, the logarithm in \eqref{eq:logint} is a sum of terms that nearly cancel, and
each term carries an absolute error of about $\epsilon$ times its own size. At a central
rank the terms are of order $n$, so the integrand carries a relative error of order
$\epsilon n$, about $10^{-10}$ at $n = 10^6$. The residuals of Table~\ref{tab:largen},
which grow with $n$, are consistent with this order.
Finally, a factor whose exponent is zero is set to one without computing its logarithm.
The reason is that this logarithm may be $-\infty$, and $0 \times (-\infty)$ is not a
number in floating point.

\subsection{Computing the difference $F(z) - F(y)$} \label{app:diff}

The factor $\{F(z) - F(y)\}^{s-r-1}$ in \eqref{eq:double}, with $F_T$ in place of $F$ in
the computation, needs $\log\{F(z) - F(y)\}$ at every inner node when $s > r + 1$. When
$y$ and $z$ both lie far in the upper tail, $F(y)$ and $F(z)$ are both near one and are
stored with an absolute error of about $\epsilon$ (Section~\ref{sec:background}). Their
difference $D = F(z) - F(y)$ then has a relative error of about $\epsilon/D$, which is
large when $D$ is small, and $(s-r-1)\log D$ inherits it.

We therefore form the difference on the side where the probabilities are small. When both
points lie below the median, $z \le 0$ in standardized units, we compute
$D = F(z)\{1 - F(y)/F(z)\}$, with the ratio obtained from the log-tails as
$F(y)/F(z) = \exp\{\log F(y) - \log F(z)\}$. When both points lie above the median,
$y \ge 0$, we compute $D = S(y)\{1 - S(z)/S(y)\}$ in the same way, where $S = 1 - F$
denotes the upper tail probability. When the two points lie on different sides of the
median, each of $F(y)$ and $F(z)$ is stored with an absolute error of about $\epsilon$ at
most, and we compute $F(z) - F(y)$ directly. In the first
two cases the difference is formed from two numbers of full relative precision. Writing
$\kappa$ for the ratio $F(y)/F(z)$ or $S(z)/S(y)$, we obtain $\log(1 - \kappa)$ from
\code{log1p()}, which is accurate when $\kappa$ is small. Precision is lost only when $z$
is close to $y$, where $\kappa$ is close to one and $1 - \kappa$ cancels. But there $D$ is
small, so the factor $D^{s-r-1}$ is small too.

We now bound the error that remains, for two points above the median. The ratio is formed
as $\kappa = \exp\{\log S(z) - \log S(y)\}$ from two stored logarithms, each with an
absolute error of about $\epsilon$ times its size, so $\kappa$ carries a relative error of
about $\epsilon |\log S(z)|$. The absolute error of $1 - \kappa$ is then about
$\epsilon \kappa |\log S(z)|$. Since $|\log S(z)| = |\log S(y)| + |\log \kappa|$ and
$\kappa |\log \kappa| \le 1/e$, this error is at most about
$\epsilon \{|\log S(y)| + 1\}$. Then $D = S(y)(1 - \kappa)$ has an absolute error of at
most about $\epsilon \{|\log S(y)| + 1\}\, S(y)$, and the power multiplies it by
$(s-r-1) D^{s-r-2}$. The absolute error of the factor is therefore at most about
$(s-r-1)\,\epsilon\,\{|\log S(y)| + 1\}\, S(y)\, D^{s-r-2}$. Since $D \le S(y)$, this error
is at most $(s-r-1)\,\epsilon\,\{|\log S(y)| + 1\}$ times $S(y)^{s-r-1}$, which is the
largest value that the factor takes for that $y$. Where $|\log S(y)|$ is large,
$S(y)^{s-r-1}$ itself is negligible. Relative to the factor itself, the error is of order
$\epsilon n$ at most near the peak of the integrand, the same order as the error of the
logarithmic scale in Appendix~\ref{app:log}. The case of two
points on different sides of the median has the same bound with
$\{|\log S(y)| + 1\}\, S(y)$ replaced by one, because $F(y)$ and $F(z)$ then carry
absolute errors of about $\epsilon$.

Two special cases remain. For adjacent ranks, $s = r + 1$, the exponent is zero, so the
factor is one and the difference is not computed. If both log-tails underflow to
$-\infty$, both probabilities lie below the smallest double, so the difference is zero in
double precision and its logarithm is set to $-\infty$.

\subsection{Closed forms, the moment condition and the internal checks} \label{app:closed}

This subsection gives four details of Section~\ref{sec:closed}: the sources of the closed
forms and the facts behind them, the derivation of the moment condition, the reason for the
unit row sums of the normal distribution, and the reasons why the error estimate of the
package is not a bound.

The uniform and exponential forms are those of
\citet[Section~3.1, eq.~2.5.5 and Exercise~3.2.1]{david2003order}. The logistic forms are
the first two cumulants of $X_{(r)}$ in \citet[eq.~2.11]{gupta1965exact}, written for the
standard logistic distribution. The Gumbel values are the moments of the extreme-value
distribution with location parameter $\log n$, by the integrals (3.2) and (3.3) of
\citet{lieblein1953exact}. The forms follow from four classical facts. For the uniform
distribution on $(0, 1)$, $X_{(r)}$ has the $\mathrm{Beta}(r, n-r+1)$ distribution. For
the exponential distribution with unit rate, $X_{(r)} = \sum_{i=1}^{r} E_i/(n-i+1)$, where
$E_i = (n-i+1)(X_{(i)} - X_{(i-1)})$, with $X_{(0)} = 0$, are independent unit exponential
variables, and $X_{(s)} - X_{(r)}$ is independent of
$X_{(r)}$. For the logistic distribution, $X_{(r)} = \log U_{(r)} - \log(1 - U_{(r)})$,
where $U_{(r)}$ has a beta distribution, whose log-moments are digamma and trigamma
values. For the Gumbel distribution in the maximum form, the maximum of $n$ observations
is again Gumbel, with location parameter $\log n$.

We now derive the decay rates of the density $f_{r:n}$ in \eqref{eq:single} and the moment
condition of Section~\ref{sec:closed}. Write
$C_{r:n} = n!/\{(r-1)!\,(n-r)!\}$ for the coefficient in \eqref{eq:single}. In the lower
tail, the condition on the parent distribution states that, for some $x_0 > 0$ and some
constants $0 < \lambda \le \Lambda$,
\begin{equation*}
\lambda |x|^{-\nu_L - 1} \le f(x) \le \Lambda |x|^{-\nu_L - 1}, \qquad x < -x_0 .
\end{equation*}
Since $\int_{-\infty}^{x} |u|^{-\nu_L - 1}\, du = |x|^{-\nu_L}/\nu_L$ for $x < 0$,
integrating these bounds gives
\begin{equation*}
\frac{\lambda}{\nu_L}\, |x|^{-\nu_L} \le F(x) \le \frac{\Lambda}{\nu_L}\, |x|^{-\nu_L},
\qquad x < -x_0 .
\end{equation*}
Since $F(x) \to 0$ as $x \to -\infty$, there is an $x_1 \le -x_0$ with $F(x) \le 1/2$ for
$x < x_1$, so that $2^{-(n-r)} \le \{1 - F(x)\}^{n-r} \le 1$ there. Multiplying the three
factors in \eqref{eq:single}, we obtain, for $x < x_1$,
\begin{equation*}
\frac{C_{r:n}\, \lambda^r}{2^{n-r}\, \nu_L^{r-1}}\, |x|^{-\nu_L r - 1}
\le f_{r:n}(x) \le
\frac{C_{r:n}\, \Lambda^r}{\nu_L^{r-1}}\, |x|^{-\nu_L r - 1},
\end{equation*}
because $\nu_L (r - 1) + \nu_L + 1 = \nu_L r + 1$. That is, $f_{r:n}$ decays like
$|x|^{-\nu_L r - 1}$ in the lower tail. The upper tail is the same with $1 - F$ in place of
$F$. There, $1 - F(x)$ lies between constant multiples of $x^{-\nu_U}$ for $x > x_0$, and
$F(x)^{r-1}$ lies between $2^{-(r-1)}$ and one beyond some point $x_2 \ge x_0$. So $f_{r:n}$ decays like
$x^{-\nu_U (n-r) - \nu_U - 1} = x^{-\nu_U (n-r+1) - 1}$.

The moment condition follows. We split $\Ex|X_{(r)}|^k = \int |x|^k f_{r:n}(x)\, dx$ at
$x_1$ and $x_2$. The three parts are nonnegative, and the middle part is at most
$\max(|x_1|, |x_2|)^k$, because $f_{r:n}$ integrates to one. So the moment is finite if and
only if both tail parts are finite. By the two-sided bounds, the lower part is finite if and only if
$\int_{-\infty}^{x_1} |x|^{k - \nu_L r - 1}\, dx$ is finite, that is, if and only if
$\nu_L r > k$. In the same way, the upper part is finite if and only if
$\nu_U (n - r + 1) > k$. A tail with the tail index $\infty$ always contributes a finite
part. The reason is that $F(x)^{r-1}\{1 - F(x)\}^{n-r} \le 1$, so
$f_{r:n}(x) \le C_{r:n} f(x)$, and $\int |x|^k f(x)\, dx$ over a bounded tail, or over a
tail that decays faster than any power (that is, $|x|^m f(x) \to 0$ for every $m$ in that
tail), is finite for every $k$. A tail of neither kind, such as a density with a logarithmic
factor like $x^{-3}\log x$, is outside the condition of Section~\ref{sec:closed}.

For the normal distribution with variance $\sigma^2$, every row of $\bB$ sums to
$\sigma^2$, at every $n$. The reason is that $X_{(r)} - \bar X$ is the $r$th order
statistic of the deviations $X_i - \bar X$, which are independent of the sample mean
$\bar X$:
\begin{equation*}
\sum_s \sigma_{r,s:n} = \Cov\Bigl(X_{(r)}, \sum_s X_{(s)}\Bigr) = n\Cov(X_{(r)}, \bar X)
= n\Cov(X_{(r)} - \bar X, \bar X) + n\Var(\bar X) = \sigma^2 .
\end{equation*}
This row-sum identity does not hold for other parent distributions in general. For example, the
exponential distribution with unit rate at $n = 2$ has row sums $1/2$ and $3/2$.

The package also reports the sum of the error estimates that \code{integrate()} returned
for the single integrals and the outer integrals. That sum is an estimate, not a bound,
for two reasons. The inner integral of a covariance does not pass its error estimate to
the outer one. And the error of a mean is not propagated into the variance that
subtracts its square, or into the covariances that subtract its product with another
mean. The evidence of accuracy in this paper is therefore the agreement with closed
forms, identities and printed tables in Section~\ref{sec:accuracy}, not this estimate.

\section{Further examples of use} \label{app:examples}

This appendix adds five examples to Section~\ref{sec:package}: a full covariance matrix
with its run-time checks, a family whose moments come from closed forms, the parameters of
a built-in family, a second user-supplied distribution, and the BLUE for a logistic sample.

\subsection{Full matrices and the run-time checks} \label{app:full}

When every rank is requested, the result is the full $n \times n$ covariance matrix. The
permutation identities \eqref{eq:identities} then apply, and so does the reflection check
for a symmetric family. A symmetric family also halves the work. The package computes
the covariances of the pairs with $r + s \le n + 1$, about half of them, and fills the
remaining pairs by the reflection $\sigma_{r,s:n} = \sigma_{n+1-s,\,n+1-r:n}$. The package
knows which built-in families are symmetric, and for a user-supplied distribution the user
declares the symmetry. The package then checks that the
means and second moments it computed, which the reflection does not touch, are
symmetric. The residuals are stored in \code{checks}:
\needspace{12\baselineskip}
\begin{CodeChunk}
\begin{CodeInput}
R> full <- os_moments(10)
R> full$checks
\end{CodeInput}
\begin{CodeOutput}
$reflection
[1] 1.110223e-16

$identities
   sum_alpha       sum_sq        sum_B     row_sums 
6.661334e-16 3.552714e-15 3.552714e-15 2.220446e-16 

\end{CodeOutput}
\end{CodeChunk}
 When the covariances have no closed form, a full matrix needs one double integral per
pair, $n(n-1)/2$ of them, or about half as many for a symmetric family. Its cost
therefore grows with the square of $n$ (Section~\ref{sec:timing}).

\subsection{Closed forms for an exponential sample} \label{app:exp}

By default the package uses a closed form wherever the family has one, and the
\code{route} field of the result reports it. For example, for twenty exponential waiting
times with mean two, the means of the minimum, the tenth order statistic, and the maximum
come from closed forms:
\needspace{8\baselineskip}
\begin{CodeChunk}
\begin{CodeInput}
R> e <- os_moments(20, ranks = c(1, 10, 20), dist = "exp", rate = 1/2)
R> e$alpha
R> e$route
\end{CodeInput}
\begin{CodeOutput}
      r1      r10      r20 
0.100000 1.337543 7.195479 
[1] "closed"
\end{CodeOutput}
\end{CodeChunk}
 
\subsection{The parameters of a built-in family} \label{app:bounded}

The parameters of a built-in family are given as in \proglang{R}, and this appendix shows
two examples. The gamma
distribution with shape two is the case that Section~\ref{sec:tables} compares with the
table of \citet{gupta1960order}. Here it has the rate three, and the field \code{standard}
holds the moments of the standard member, the gamma distribution with shape two and rate
one. The means returned are one third of the means of the standard member:
\needspace{8\baselineskip}
\begin{CodeChunk}
\begin{CodeInput}
R> g <- os_moments(10, ranks = c(1, 5, 10), dist = "gamma", shape = 2, rate = 3)
R> g$alpha
R> g$standard$alpha / 3
\end{CodeInput}
\begin{CodeOutput}
       r1        r5       r10 
0.1553405 0.5250292 1.5409857 
       r1        r5       r10 
0.1553405 0.5250292 1.5409857 
\end{CodeOutput}
\end{CodeChunk}
 Student's $t$ distribution has only its degrees of freedom as a parameter, so a shift and a
multiplier are given by \code{os_shift} and \code{os_multiplier}. For $Y = 10 + 2X$, where
$X$ has Student's $t$ distribution with five degrees of freedom:
\needspace{8\baselineskip}
\begin{CodeChunk}
\begin{CodeInput}
R> t5 <- os_moments(5, dist = "t", df = 5, os_shift = 10, os_multiplier = 2)
R> t5$alpha
R> t5$var
\end{CodeInput}
\begin{CodeOutput}
       r1        r2        r3        r4        r5 
 7.098050  8.883987 10.000000 11.116013 12.901950 
      r1       r2       r3       r4       r5 
4.591509 1.708025 1.400664 1.708025 4.591509 
\end{CodeOutput}
\end{CodeChunk}
 The heavy tails of the distribution show in the result: the variance of the sample minimum
is more than three times the variance of the sample median.

\subsection{A symmetric user-supplied distribution with a kink} \label{app:laplace}

The Laplace distribution, with the density $e^{-|x|}/2$, is a second example of a
user-supplied distribution (Section~\ref{sec:userdist}). It shows two more arguments of
\code{os_dist()}. First, the distribution is symmetric about its median,
and the user declares this with \code{symmetric = TRUE}, which halves the double integrals
of a full matrix and turns on the reflection check (Appendix~\ref{app:full}). Second, the
density has a kink at the median, where it is not differentiable, and the user declares
this point with \code{breaks = 0}. The package then cuts every integral at the declared
point (Appendix~\ref{app:split}). The
distribution function below takes the upper tail from the lower tail by the symmetry:
\needspace{12\baselineskip}
\begin{CodeChunk}
\begin{CodeInput}
R> laplace <- os_dist(
+    p = function(q, lower.tail = TRUE, log.p = FALSE) {
+      s <- if (lower.tail) q else -q
+      lp <- ifelse(s <= 0, log(0.5) + s, log1p(-0.5 * exp(-abs(s))))
+      if (log.p) lp else exp(lp)
+    },
+    d = function(x, log = FALSE) if (log) log(0.5) - abs(x) else 0.5 * exp(-abs(x)),
+    q = function(p) ifelse(p <= 0.5, log(2 * p), -log(2 * (1 - p))),
+    symmetric = TRUE, breaks = 0, name = "laplace")
R> os_moments(5, dist = laplace)
\end{CodeInput}
\begin{CodeOutput}
Exact moments of order statistics: laplace, n = 5, ranks 1 2 3 4 5
alpha (means):
           r1            r2            r3            r4            r5 
-1.588542e+00 -5.729167e-01 -3.206503e-17  5.729167e-01  1.588542e+00 
B (covariance matrix):
          r1        r2        r3        r4        r5
r1 1.4702510 0.4825716 0.2640278 0.1994770 0.1847146
r2 0.4825716 0.5024609 0.2801736 0.2144835 0.1994770
r3 0.2640278 0.2801736 0.3511806 0.2801736 0.2640278
r4 0.1994770 0.2144835 0.2801736 0.5024609 0.4825716
r5 0.1847146 0.1994770 0.2640278 0.4825716 1.4702510
route: integrate; quadrature error estimate: 4.82e-10
checks: reflection 4.44e-16; identities sum_alpha 5.37e-17,
  sum_sq 0.00e+00, sum_B 1.78e-15
\end{CodeOutput}
\end{CodeChunk}
 In the last line of the output, the residual \code{sum_B} shows that the covariances sum
to their exact value, $n \Var(X) = \rpLaplaceSumBExact$, up to rounding. As a further check, we computed each covariance by a
double integral in $x$ whose range is cut at the kink, and the ten covariances of the
output agree with these integrals within $10^{-12}$.

The argument \code{breaks} matters for the covariances. When we omit it from the call
above, the sum of the covariances differs from $\rpLaplaceSumBExact$ by
$\rpLaplaceNoBreaksSumB$, and the covariance of the ranks \rpLaplaceNoBreaksPair{} differs
from its double integral by $\rpLaplaceNoBreaksDiff$. This difference is larger than the
error estimate that the package reports for the whole computation, so the error estimate
does not reveal the loss. The residual \code{sum_B} of the run-time checks does reveal it, when the rank
set is complete. The means and the variances change far less. A point where the density
jumps is declared in the same way. The package cannot find such points itself, and the
built-in families have none.

\subsection{The BLUE for a logistic sample} \label{app:logistic}

The same call serves any location-scale family and any subset of ranks. For a complete
logistic sample of $n = \rpLogisN$:
\needspace{10\baselineskip}
\begin{CodeChunk}
\begin{CodeInput}
R> bl <- os_blue(os_moments(9, dist = "logis"))
R> round(bl$w_mu, 4)
R> round(bl$w_sigma, 4)
R> c(var_mu = bl$var_mu, var_sigma = bl$var_sigma, cov = bl$cov)
\end{CodeInput}
\begin{CodeOutput}
    r1     r2     r3     r4     r5     r6     r7     r8     r9 
0.0320 0.0901 0.1335 0.1600 0.1689 0.1600 0.1335 0.0901 0.0320 
     r1      r2      r3      r4      r5      r6      r7      r8      r9 
-0.0973 -0.0946 -0.0714 -0.0380  0.0000  0.0380  0.0714  0.0946  0.0973 
      var_mu    var_sigma          cov 
3.424736e-01 8.626559e-02 1.412826e-17 
\end{CodeOutput}
\end{CodeChunk}
 The covariance of the two estimators is zero. The reason is that the standard logistic
distribution is symmetric about zero and the complete rank set contains $n + 1 - r$ with
every rank $r$. So $\balpha_Z$ is antisymmetric, and the quantity $\bm
1'\bB_Z^{-1}\balpha_Z$, to which the covariance is proportional, is zero. The estimated
scale $\sigma$ is the logistic scale parameter, and the standard deviation of the parent
distribution is \rpLogisSdPerScale{} times this scale parameter. The variances of the two
estimators are $\Var(\hat\mu) = \rpLogisVarLoc\,\sigma^2$ and $\Var(\hat\sigma) =
\rpLogisVarScale\,\sigma^2$.

\section{Further accuracy results} \label{app:accuracy}

This appendix gives the further results behind the three layers of evidence in
Section~\ref{sec:accuracy}: the closed forms, the identities on full matrices, and the
printed tables with the cells on which they disagree with the package.

\subsection{Closed forms} \label{app:closedforms}

Table~\ref{tab:closed} gives the largest residual of each family in three ranges of $n$, for
the comparison of Section~\ref{sec:closedforms}. The uniform and the exponential
distributions are checked at every mean, variance, and covariance of the five-number ranks,
\atClosedCovEntries{} covariance entries in all. The logistic distribution is checked at the
means and variances, and the Gumbel distribution at the mean and variance of the sample
maximum. From $2 \le n \le 20$ to $101 \le n \le 1001$, the largest residual of a family
\atTrendClosed. The package's own error estimate is at most \atClosedQuadErrMax{} over these
runs, and \atQuadErrRelation.

At $n = 2$ the covariance has an exact value for any parent distribution with a mean, which
gives a check for the logistic and the Gumbel distributions, whose covariances have no
closed form. Since $X_{(1)}X_{(2)} = X_1X_2$ and $\Ex[X_1X_2] = (\Ex X)^2$, we have
$\sigma_{1,2:2} = (\Ex X)^2 - \mu_{1:2}\mu_{2:2}$. The quadrature differs from this
$\sigma_{1,2:2}$ by \atAnchorLogis{} for the logistic distribution and by \atAnchorGumbel{}
for the Gumbel distribution.

\begin{table}[t!]
\centering
\footnotesize
\setlength{\tabcolsep}{4pt}
\begin{tabular}{llrrr}
\toprule
family & quantities checked & $2 \le n \le 20$ & $21 \le n \le 100$ & $101 \le n \le 1001$ \\
\midrule
uniform & means, variances, covariances & $2.22 \times 10^{-16}$ & $2.22 \times 10^{-16}$ & $2.22 \times 10^{-16}$ \\
exponential & means, variances, covariances & $8.88 \times 10^{-16}$ & $2.00 \times 10^{-15}$ & $5.33 \times 10^{-15}$ \\
logistic & means, variances & $1.55 \times 10^{-15}$ & $1.78 \times 10^{-15}$ & $2.00 \times 10^{-15}$ \\
Gumbel & mean and variance of the maximum & $8.88 \times 10^{-16}$ & $2.00 \times 10^{-15}$ & $2.22 \times 10^{-15}$ \\
\bottomrule
\end{tabular}
 \caption{\label{tab:closed} The largest absolute difference between the quadrature and
the closed form, by family and range of $n$, over \atClosedSizes{} sample sizes at the
five-number ranks and their pairs.}
\end{table}

Table~\ref{tab:largen} gives the residuals of the comparison beyond
$n = \atClosedSizeMax$, for the uniform and the exponential distributions at \atLargeNEq.
The relative error of a covariance is about its correlation residual divided by its
correlation, up to the relative errors of the two variances, which are below
\atLargeNVarWorst{} here. So a pair of order statistics that is nearly independent keeps
fewer digits of its covariance.

\begin{table}[t!]
\centering
\footnotesize
\setlength{\tabcolsep}{3.5pt}
\begin{tabular}{llrrrrr}
\toprule
family & $n$ & \shortstack[r]{means\\(relative)} & \shortstack[r]{variances\\(relative)} & \shortstack[r]{correlations\\(absolute)} & \shortstack[r]{extremes'\\correlation (exact)} & \shortstack[r]{extremes' covariance\\(relative)} \\
\midrule
uniform & $10^{4}$ & $5.55 \times 10^{-13}$ & $6.42 \times 10^{-13}$ & $5.69 \times 10^{-13}$ & $1.00 \times 10^{-4}$ & $6.61 \times 10^{-11}$ \\
uniform & $10^{5}$ & $5.55 \times 10^{-12}$ & $3.76 \times 10^{-12}$ & $3.76 \times 10^{-12}$ & $1.00 \times 10^{-5}$ & $2.37 \times 10^{-9}$ \\
uniform & $10^{6}$ & $5.55 \times 10^{-11}$ & $7.99 \times 10^{-11}$ & $2.57 \times 10^{-11}$ & $1.00 \times 10^{-6}$ & $5.62 \times 10^{-8}$ \\
\midrule
exponential & $10^{4}$ & $2.22 \times 10^{-15}$ & $8.01 \times 10^{-13}$ & $5.02 \times 10^{-13}$ & $7.80 \times 10^{-5}$ & $4.38 \times 10^{-11}$ \\
exponential & $10^{5}$ & $1.22 \times 10^{-14}$ & $4.36 \times 10^{-12}$ & $2.76 \times 10^{-12}$ & $7.80 \times 10^{-6}$ & $3.54 \times 10^{-7}$ \\
exponential & $10^{6}$ & $1.11 \times 10^{-10}$ & $7.70 \times 10^{-11}$ & $3.47 \times 10^{-11}$ & $7.80 \times 10^{-7}$ & $8.58 \times 10^{-6}$ \\
\bottomrule
\end{tabular}
 \caption{\label{tab:largen} The residuals at \atLargeNEq, at the five-number ranks of the
uniform and the exponential distributions: the largest relative residual of the means and
of the variances, and the largest residual of the correlation $\rho_{rs}$ over the ten
pairs. The last two columns concern the pair of the minimum and the maximum (the
extremes): their exact correlation and the relative residual of their covariance.}
\end{table}

\subsection{Identities on full matrices} \label{app:identities}

Table~\ref{tab:ident} gives the largest residual of each identity by family, for the
comparison of Section~\ref{sec:identities}. The matrices of the normal distribution are
those at every $n$ from 2 to 100, \atIdentSizesNorm{} matrices in all. The reflection
residual is at most \atIdentReflWorst{} over the symmetric families.

\begin{table}[t!]
\centering
\footnotesize
\setlength{\tabcolsep}{4pt}
\begin{tabular}{llrrrrr}
\toprule
family & sizes $n$ & $\sum_r \alpha_r$ & $\sum_r E[X_{(r)}^2]$ & $\sum_{r,s} B_{rs}$ & row sums of $B$ & reflection \\
\midrule
normal & 2 to 100 (99) & $1.02 \times 10^{-14}$ & $5.68 \times 10^{-14}$ & $7.18 \times 10^{-13}$ & $1.93 \times 10^{-13}$ & $2.66 \times 10^{-15}$ \\
uniform & 5, 10, 25, 50, 100 & $3.55 \times 10^{-15}$ & $2.13 \times 10^{-14}$ & $3.98 \times 10^{-13}$ & -- & $4.44 \times 10^{-16}$ \\
exponential & 5, 10, 25, 50, 100 & $3.55 \times 10^{-15}$ & $1.42 \times 10^{-14}$ & $9.95 \times 10^{-13}$ & -- & -- \\
logistic & 5, 10, 25, 50, 100 & $1.60 \times 10^{-14}$ & $1.42 \times 10^{-14}$ & $4.41 \times 10^{-13}$ & -- & $3.55 \times 10^{-15}$ \\
Gumbel & 5, 10, 25, 50, 100 & $4.97 \times 10^{-14}$ & $2.84 \times 10^{-14}$ & $5.83 \times 10^{-13}$ & -- & -- \\
\bottomrule
\end{tabular}
 \caption{\label{tab:ident} The largest residual of each identity in \eqref{eq:identities}
on full matrices computed by quadrature, by family: the absolute difference between the two
sides of the identity, in the standardized units in \eqref{eq:units}, against a tolerance of
$10^{-9}$. The unit row sums are checked for the normal distribution only, where they hold
at every $n$, and the reflection residual exists for the symmetric families only, with a
dash where neither applies.}
\end{table}

For the normal distribution, the identities also carry the evidence at the top of the range,
where no printed table reaches. At $n = \atLargeN$, the covariance matrix has
\atLargeEntries{} entries below the diagonal. Of these, \atLargePairs{} were computed as
pairs, each by one call of \code{os_moments()} at two ranks. The rest were filled by the
reflection $\sigma_{r,s:n} = \sigma_{n+1-s,\,n+1-r:n}$, so each row sum tests the computed
entries together with the reflected ones. Every one of the \atLargeN{} row sums lies within
\atLargeRowMax{} of one, with a mean absolute deviation of \atLargeRowMean. The sum of all
entries is within \atLargeTotal{} of $n$. The identities on the first and second moments
hold within \atLargeSumAlpha{} and \atLargeSumSq.

\subsection{The printed tables in the literature} \label{app:tables}

This subsection describes the conditional-beta check of Section~\ref{sec:tables} and
discusses the cells on which a printed table disagrees with the package.

The conditional-beta check works in the probability variable $u = F(x)$, in which
$X_{(r)} = Q(U_{(r)})$ with $U_{(r)} \sim \mathrm{Beta}(r, n-r+1)$
(Section~\ref{sec:background}). It computes the first moment of $X_{(r)}$ as
\begin{equation*}
\mu_{r:n} = \frac{1}{\mathrm{B}(r, n-r+1)} \int_0^1 Q(u)\, u^{r-1} (1-u)^{n-r}\, du ,
\end{equation*}
which is the integral in \eqref{eq:moments} after the substitution $u = F(x)$, and the
second moment by the same integral with $Q(u)^2$ in place of $Q(u)$. For a product moment
with $r < s$, it uses the representation $U_{(s)} = U_{(r)} + (1 - U_{(r)})\,V$, where $V
\sim \mathrm{Beta}(s - r, n - s + 1)$ is independent of $U_{(r)}$. The single integrals, and
the double integral of a product moment over the two beta variables, are computed by
\code{integrate()} against beta densities. A disagreeing cell is attributed to the table
when the check agrees with the package within $10^{-10}$. It agreed within
\cpRouteMaxDiffAll{} on every one of the \cpCellsTable{} cells. The counts leave out
\countword{\cpTietjenDoubtful} cell of the table of \citet{tietjen1976variances} that is
illegible in the scan we worked from.

We split the \cpCellsTable{} disagreements by size. A cell more than ten units of its
criterion away from the package is a gross error of the table. A cell within ten units
is a small error, of the size of the table's rounding or its stated accuracy. Two tables
carry gross errors. Of the \cpTietjenFail{} cells of \citet{tietjen1976variances} that
disagree, \cpTietjenGross{} lie at the quartile ranks of $n \ge \cpTietjenFailNmin$.
They are off by up to \cpTietjenWorstFail, which is \cpTietjenUnitsMax{} units of
$10^{-7}$, far beyond the seven decimals the table certifies. The other \cpTietjenFine{}
lie within \cpTietjenFineUnitsMax{} units of $10^{-7}$, beyond the certified accuracy
but by less than ten units of it. The authors describe a fixed 128-point Gauss--Hermite
rule for their outer integral. Such a rule may resolve the narrowing integrands of the
central ranks less well as $n$ grows. This is a possible explanation of the pattern, and
we did not rerun their implementation to confirm it.

The table of \citet{gupta1960order} for the gamma distribution with shape two is the other
table with gross errors. We compared it through \code{os_dist()}, with the gamma
distribution supplied by its three functions. At $n = 5$ it agrees on \countword{\cpGuptaFiveAgree} of its ten cells,
with \countword{\cpGuptaFiveFail} cell that is \cpGuptaFiveUnits{} units of its sixth
significant figure away from the package's value. At $n = 10$ it disagrees on
\cpGuptaTenFail{} of \cpGuptaTenCells{} cells, all at ranks \cpGuptaTenFailRankMin{} and
above, by \cpGuptaTenUnitsMin{} to \cpExactTenUnitsMax{} units of the sixth significant
figure. For this family, a third way of computing the moments needs no floating point at
all. With $F(x) = 1 - (1 + x)e^{-x}$, the binomial expansions of $F(x)^{k-1}$ and of $(1 +
x)^m$ turn every moment into a finite sum of rationals, which we summed in exact rational
arithmetic. The rational values agree with the package within \cpExactPkgMax{} on all
\cpExactCells{} cells. They confirm the \cpExactTenFail{} printed errors at $n = 10$, up to
\cpExactTenMax. These alternating sums are Gupta's own formulas (2.7) to (2.9), which he
computed numerically in 1960. A loss of precision by cancellation in these sums is therefore
a consistent explanation of his errors, although we did not rerun his arithmetic.

The other disagreements are small. The \cpCellsFine{} cells within ten units of their
criterion lie within \cpFineUnitsMax{} units of it. For the sources on the normal
distribution other than the table of \citet{tietjen1976variances}, these cells lie within
\cpOtherUnitsMax{} units of the tenth decimal, which is the rounding of those tables. For
the table of \citet{gupta1960order}, these cells are the smallest of the errors described
above.
Table~\ref{tab:tables} gives the largest difference of each source among its agreeing
cells. The extreme-value example of \citet{lieblein1953exact} agrees within
\cpLiebleinWorstMean{} on the means and \cpLiebleinWorstCov{} on the variances and
covariances, which he gives to about four decimals. The \countword{\cpLogisCells} logistic cells
converted from the exact expressions of \citet{gupta1965exact} agree within
\cpLogisWorst. We found no numeric table of logistic covariances. Beyond the $n = 2$
check of Appendix~\ref{app:closedforms}, the logistic covariances therefore rest on the
identities.

\subsection{The other built-in families} \label{app:newfamilies}

This appendix gives the details of Section~\ref{sec:newfamilies}: the reference, the checks
of the eight families at ordinary parameter values, the two harder parameter regions, and
what \pkg{SciPy} returns in the same cases.

\paragraph{The reference.} The reference for one order statistic is the moment
$\Ex[\{Q(U_{(r)}) - c_r\}^k]$, where $U_{(r)} \sim \mathrm{Beta}(r, n - r + 1)$ and $k = 1, 2$,
integrated in the probability variable $u$. It uses the quantile function only. The
package's integrands use $F$ and $f$, so the two computations share no integrand. They do
share \code{integrate()}, the center $c_r$, and \proglang{R}'s implementation of each family.
We cut the range of
$u$ at one half and take the quantile of each half from its own tail. In the tail
probability $w$ of a half ($w = u$ in the lower half and $w = 1 - u$ in the upper half), we
cut at $p \pm 10 s$ and at $p + 10 \cdot 4^{j} s$ for $j = 1, 2, \dots$, where $p$ and $s$
are the mean and the standard deviation of the beta density in $w$. Where $|Q|^k$ grows like
$w^{-g}$ in a tail, as for Student's $t$ distribution with $g = k/\nu$, the integrand is of
the order of $w^{r - 1 - g}$ in the lower tail, and we integrate in $w^{r - g}$, in which it is
bounded. Such an exponent exists if and only if $r > g$, which is the condition for the
moment to exist in that tail. We checked the reference against the two closed forms that
these families have. For the Weibull distribution with shape $c$ and
$F(x) = 1 - \exp(-x^c)$, expanding $F^{r-1}$ in
\eqref{eq:single} gives
\begin{equation*}
\Ex[X_{(r)}^k] = r \binom{n}{r}\, \Gamma\Bigl(1 + \frac{k}{c}\Bigr)
\sum_{j=0}^{r-1} (-1)^j \binom{r-1}{j} (n - r + 1 + j)^{-(1 + k/c)} ,
\end{equation*}
which we use at the ranks 1 and 2. For Student's $t$ distribution with two degrees of
freedom, $Q(u) = (2u - 1)/\{2u(1 - u)\}^{1/2}$, so
\begin{equation*}
\Ex[X_{(r)}] = \frac{2\mathrm{B}(r + \tfrac12, n - r + \tfrac12) - \mathrm{B}(r - \tfrac12, n - r +
\tfrac12)}{\sqrt2\, \mathrm{B}(r, n - r + 1)}, \qquad
\Ex[X_{(r)}^2] = \frac{n(n-1)}{2(r-1)(n-r)} - 2 ,
\end{equation*}
where the formula for $\Ex[X_{(r)}]$ holds at every rank, and the formula for
$\Ex[X_{(r)}^2]$ holds for $2 \le r \le n - 1$, the ranks at which the second moment exists. Over
\fmRefValidCells{} order statistics of these two
distributions at sample sizes up to 1001, the reference agrees with the closed forms within
$\fmRefValidWorstMean$ on the means and $\fmRefValidWorstVar$ on the variances, in
relative terms.

\paragraph{The eight families.} Table~\ref{tab:newfamilies} gives the checks by family, at
\fmCheckCases{} parameter settings. We apply three checks. First, the package agrees with
the reference within $\fmCheckPkgRefWorst$ on every mean and variance. Here a variance is
compared in relative terms, and a mean relative to the larger of its absolute value and the
standard deviation of the order statistic, so that a mean of zero is judged on the scale of
its own order statistic. Second, the permutation identities \eqref{eq:identities} hold within
$\fmIdentWorst$ in relative terms on \fmIdentMatrices{} full matrices at $n = 10$ and 25.
Third, the covariances between the five-number ranks at $n = \fmCovN$ agree with the
conditional-beta check of Appendix~\ref{app:tables} on \fmCovPairs{} pairs of ranks, where
the largest absolute difference between the two correlations is $\fmCovWorst$. For the
other \fmCovRouteFailed{} pairs, the nested integral of the check itself does not converge.

On the same means and variances, \pkg{SciPy} agrees with the reference within $\fmTolScipy$
except on \fmCheckScipyOver{} order statistics, \fmCheckScipyOverWhere. The largest
difference of \pkg{SciPy} is \fmCheckScipyWorstPct\%, at rank \fmCheckScipyWorstR{} of
$n = \fmCheckScipyWorstN$ observations for \fmCheckScipyWorstCase.

\begin{table}[t!]
\centering
\footnotesize
\setlength{\tabcolsep}{3.5pt}
\begin{tabular}{l>{\raggedright\arraybackslash}p{4.3cm}rrrrr}
\toprule
Family & Parameter values & Values & Package & \pkg{SciPy} & Identities & Covariances \\
\midrule
Student $t$ & \code{df} $= 3,\ 5,\ 30$ & 150 & $1.2 \times 10^{-13}$ & $8.1 \times 10^{-13}$ & $2.5 \times 10^{-12}$ & $7.8 \times 10^{-11}$ \\
Cauchy & -- & 26 & $3.4 \times 10^{-14}$ & $8.6 \times 10^{-13}$ & -- & $1.0 \times 10^{-15}$ \\
gamma & \code{shape} $= 0.5,\ 2,\ 10$ & 150 & $1.1 \times 10^{-13}$ & $2.5 \times 10^{-6}$ & $1.3 \times 10^{-15}$ & $5.5 \times 10^{-12}$ \\
chi-square & \code{df} $= 1,\ 4$ & 100 & $6.8 \times 10^{-14}$ & $1.7 \times 10^{-6}$ & $6.7 \times 10^{-16}$ & $1.9 \times 10^{-12}$ \\
beta & (\code{shape1}, \code{shape2}) $= (0.5, 0.5),\ (2, 3),\ (5, 2)$ & 150 & $6.2 \times 10^{-15}$ & $2.0 \times 10^{-1}$ & $8.9 \times 10^{-16}$ & $6.4 \times 10^{-11}$ \\
lognormal & \code{sdlog} $= 0.5,\ 1$ & 100 & $1.5 \times 10^{-13}$ & $8.2 \times 10^{-13}$ & $4.4 \times 10^{-16}$ & $1.8 \times 10^{-10}$ \\
Weibull & \code{shape} $= 0.5,\ 1.5,\ 3$ & 150 & $3.1 \times 10^{-14}$ & $2.2 \times 10^{-6}$ & $3.3 \times 10^{-14}$ & $4.9 \times 10^{-11}$ \\
$F$ & (\code{df1}, \code{df2}) $= (4, 10),\ (10, 20)$ & 100 & $1.2 \times 10^{-14}$ & $8.3 \times 10^{-13}$ & $8.9 \times 10^{-16}$ & $3.1 \times 10^{-10}$ \\
\bottomrule
\end{tabular}
 \caption{\label{tab:newfamilies} The checks of the eight other built-in families. Values: the number of means and
variances compared, at the five-number ranks of $n = 5$, 25, 101, 501, and 1001 (a rank
whose variance does not exist is left out). Package and \pkg{SciPy}: the largest relative
difference from the reference integral in the probability variable, where a mean is
compared on the scale of the larger of its absolute value and the standard deviation of the
order statistic. Identities: the largest relative residual of the sum of the means and of the sum
of the covariances on full matrices at $n = 10$ and 25, against the exact moments of the
parent distribution (not applicable to the Cauchy distribution, which has no mean).
Covariances: the largest difference from the conditional-beta check between the five-number
ranks at $n = \fmCovN$, in units of the correlation.}
\end{table}

\paragraph{Two harder parameter regions.} The first region is a density that is infinite
at an endpoint: the gamma, chi-square, Weibull, beta, and $F$ distributions at indices
$\theta$ from 0.05 to 0.8 (Appendix~\ref{app:endpoint}). The second region is a variance that barely
exists: Student's $t$ distribution with 2.001 to 3 degrees of freedom at the minimum, with
1.005 to 1.5 degrees of freedom at the second order statistic, and the $F$ distribution
with $\nu_2$ from 4.02 to 6 at the maximum. Both regions are run at the five-number ranks
of $n = 5$, 25, 101, 1001, $10^4 + 1$, and $10^6 + 1$. Of the \fmRegionCells{} order
statistics in \fmRegionCases{} parameter settings, the package returns the mean and the
variance of \fmRegionReturned{}, each within $\fmRegionPkgWorst$ of the reference, and it
stops with an error for the remaining \fmRegionErrors{} (\fmRegionErrorList). \pkg{SciPy} returns a value that is
not finite for \fmRegionScipyNonFinite{} of them, at sample sizes from
$n = \fmRegionScipyNonFiniteFrom$ on, and it differs from the reference by more than
$\fmTolScipy$ on \fmRegionScipyOver{} of the other \fmRegionScipyFinite. The package's tests
add exact values where they exist. For the beta distribution with the shape parameters
$\theta$ and 1, which is the distribution of the power $1/\theta$ of a uniform variable, the
minimum has $\Ex[X_{(1)}^k] = \prod_{j=1}^{k/\theta} j/(n + j)$ when $k/\theta$ is an integer.
The package agrees
with these moments for $k = 1, 2$ within $10^{-10}$ for the mean and $10^{-9}$ for the
variance, in relative terms, at $\theta = 0.5$, 0.1, and 0.05 and $n = 10^6$.

On full matrices at $n = \fmMatN$ in the two regions, \fmMatComputed{} of
\fmMatCases{} parameter settings are computed, with the sum of the covariances within
$\fmMatWorstSumB$ of its exact value in relative terms, \fmMatErrors{} stop with an
error (\fmMatErrorCases), and the other \fmMatNotApplicable{} have a variance that does
not exist. In each of these, every covariance exists, but a piece of the range
of one of them is not resolved to the tolerance, and the package stops rather than return a
number.

\paragraph{A moment that does not exist.} \pkg{SciPy} \fmScipyVersion{} returns a finite
variance for each of the \fmScipyNonexistCells{} order statistics that we tried whose
variance does not exist (the ranks 1, 2, 4, and 5 of five Cauchy observations, and the
ranks 1 and 5 of five observations from Student's $t$ distribution with two degrees of
freedom), \fmScipyNonexistWarnPhrase. For the second of five Cauchy
observations, for example, it returns the variance \fmScipyCauchySecondVar. The package
refuses each of them.

\section{Details of the comparison and the computing time} \label{app:comparison}

This appendix gives the details behind Section~\ref{sec:comparison}. For the comparison
with other software, it gives the versions of the tools, the sample sizes, the conventions,
and the tools that we could not run. For the computing time, it gives the table of the
elapsed times.

\subsection{The comparison with other software} \label{app:tools}

The \proglang{R} packages are \cpToolVersions, and \pkg{SciPy} is version
\cpScipyVersion. The package side is \code{os_moments()} at its defaults. \pkg{EnvStats},
\pkg{SuppDists}, and \pkg{metaBLUE} ran at $n = 4m + 1$ from 5 to 101 and then at 121,
141, 161, 181, 201, 253, 301, 353, 401, 501, 601, 701, 801, and 1001, 39 sample sizes in
all. \pkg{lmomco}, the closed forms of \pkg{mos}, and \pkg{SciPy} ran at
$n = 5, 25, 101, 501, 1001$, and the Monte Carlo moments of \pkg{mos} at
$n = 5, 25, 101$. Every difference is an absolute difference in the units of the
standard member. A relative difference is in percent of the package's entry for a
variance or a covariance. For a mean, it is in percent of the largest mean of the call,
so that a mean near zero does not inflate it.

We first discuss the tools that integrate or use a closed form. \pkg{EnvStats} agrees
with the package on the normal means within \cpEnvStatsWorst{} at every sample size up
to $n = \cpEnvStatsSoundTo$. \pkg{lmomco} integrates at \proglang{R}'s default tolerance.
It agrees within \cpLmomcoSoundWorst{} up to $n = \cpLmomcoSoundTo$ on the normal,
exponential, and Gumbel means, and loses its accuracy from $n = \cpLmomcoFirstFail$ on.
For the normal distribution, \pkg{lmomco} \cpLmomcoNote. The closed forms of \pkg{mos}
for the uniform distribution agree within \cpMosUnifWorst{} at every sample size. Those for the
exponential distribution agree within \cpMosExpSoundWorst{} up to
$n = \cpMosExpSoundTo$. They then return values off by as much as \cpMosExactWorst{} at
$n = \cpMosExpFirstFail$, and non-finite values at the two larger sample sizes.
Table~\ref{tab:tools} counts the values that a tool returns after it breaks down as the
end of its range, not as errors. The function \code{order\_statistic} of \pkg{SciPy}
agrees within \cpScipyWorstMean{} on the means and \cpScipyWorstVar{} on the variances,
over \cpScipyEntries{} entries of five families.

The approximate tools differ from the package by the error of their methods.
\pkg{SuppDists} computes the normal means by the approximate form of Royston's
algorithm, documented to four decimals, and differs from the package by up to
\cpSuppDistsWorst. The series of \pkg{metaBLUE} misses the entries of $\balpha$ by up to
\cpMetaBlueAlphaRel\% and those of $\bB$ by up to \cpMetaBlueCovRel\%. Its variances of
the minimum and the maximum are off by \cpMetaBlueExtVarRelAtFive\% at $n = 5$,
\cpMetaBlueExtVarRelAtHundredOne\% at $n = 101$, and \cpMetaBlueExtVarRelAtThousand\% at
$n = 1001$. The Monte Carlo moments of \pkg{mos} for the normal distribution, at its
documented $10^5$ replicates, differ from the package by up to \cpMosMcMeanRel\% on the
means and \cpMosMcVarRel\% on the variances.

Three tools named in the literature could not be run and are left out. NAG's
implementation of Algorithm AS~128 is commercial. The \proglang{Maple} procedures of
\citet{childs2002series} need \proglang{Maple}. \proglang{Mathematica}'s
\code{OrderDistribution} gives the joint distribution of order statistics, with moments by
integration, but \proglang{Mathematica} is commercial software for which we have no
license.

\subsection{Computing time} \label{app:timing}

\begin{table}[t!]
\centering
\footnotesize
\setlength{\tabcolsep}{5pt}
\begin{tabular}{rrrrrr}
\toprule
$n$ & \shortstack[r]{singles only\\(ms)} & \shortstack[r]{with covariances\\(ms)} & \shortstack[r]{share of the\\doubles (\%)} & \shortstack[r]{per single\\(ms)} & \shortstack[r]{per double\\(ms)} \\
\midrule
5 & 3.9 & 265.0 & 99 & 0.26 & 43.5 \\
25 & 3.7 & 332.0 & 99 & 0.25 & 54.7 \\
101 & 4.5 & 484.0 & 99 & 0.30 & 79.9 \\
501 & 4.5 & 595.0 & 99 & 0.30 & 98.4 \\
1001 & 4.3 & 570.0 & 99 & 0.28 & 94.3 \\
\bottomrule
\end{tabular}
 \caption{\label{tab:timing} The elapsed time of two calls of \code{os\_moments()} for the
standard normal distribution at the five-number ranks, the median of
\countword{\atTimeReps} repetitions on one core. The first call has \code{cov = FALSE} and
computes \atTimeSingles{} single integrals. The second call has \code{cov = TRUE} and
computes \atTimeDoublesPkg{} double integrals in addition, the other pairs being filled by
reflection. The share of the double integrals is the difference of the two times divided
by the second time. The cost per single integral is the first time divided by
\atTimeSingles, and the cost per double integral is the difference of the two times
divided by \atTimeDoublesPkg.}
\end{table}

\begin{figure}[t!]
\centering
\includegraphics[width=0.75\textwidth]{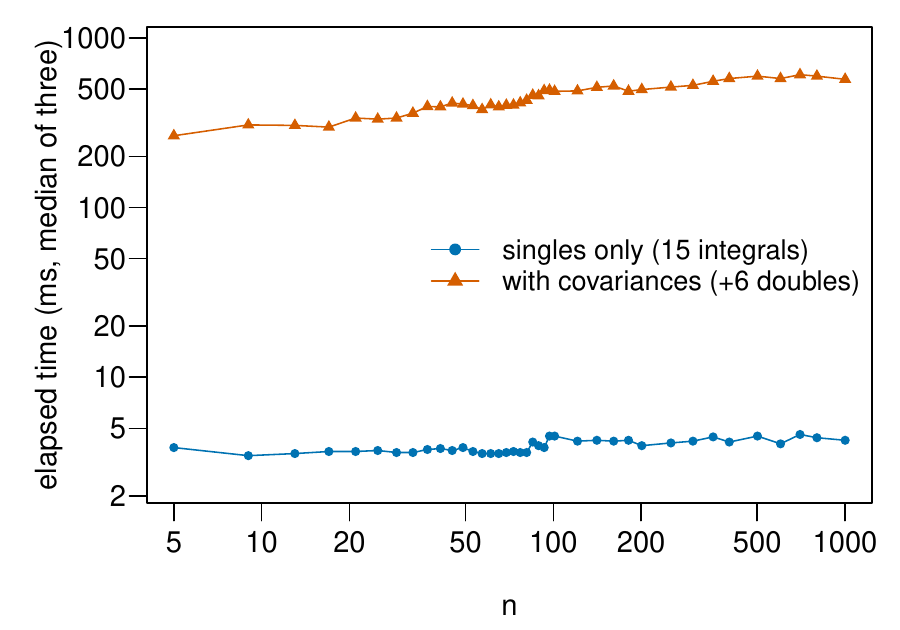}
\caption{\label{fig:timing} The elapsed time of the two calls of Table~\ref{tab:timing}
against the sample size $n$, on logarithmic axes. The lower curve is the call with
\code{cov = FALSE}, and the upper curve is the call with \code{cov = TRUE}.}
\end{figure}

Table~\ref{tab:timing} gives the elapsed times of Section~\ref{sec:timing} at five sample
sizes, and Figure~\ref{fig:timing} shows them at every sample size run from $n =
\atTimeNmin$ to $n = \atTimeNmax$. Each time is the median of \countword{\atTimeReps}
repetitions on one core of an \atTimeMachine. At the larger sample sizes of
Section~\ref{sec:closedforms}, \atLargeNEq, a call for the five ranks and their ten
covariances takes \atLargeNTimeCovMin{} to \atLargeNTimeCovMax{} seconds. This time
corresponds to about \atLargeNPerDoubleMs{} ms per double integral at $n = \atLargeNMax$,
against \atTimePerDoubleAtMax{} ms at $n = \atTimeNmax$. A possible reason why the cost
of one integral hardly depends on $n$ is the split \eqref{eq:split}, which keeps the shape
of the integrand relative to its middle piece the same at every sample size, although we
did not test this explanation. \end{appendix}
\fi

\end{document}